\documentclass[aps,prc,preprint,groupedaddress,nofootinbib]{revtex4-1}

\usepackage{graphicx}
\usepackage{amsmath,bm,ascmac,amssymb,calc}

\begin{document}


\title{Exposing minimal composition of Kohn-Sham theory
and its extendability}


\author{H. Nakada}
\email[E-mail:\,\,]{nakada@faculty.chiba-u.jp}

\affiliation{Department of Physics, Graduate School of Science,
 Chiba University,\\
Yayoi-cho 1-33, Inage, Chiba 263-8522, Japan}


\date{\today}

\begin{abstract}
  Reducing the many-fermion problem
  to a set of single-particle (s.p.) equations,
  the Kohn-Sham (KS) theory has provided a practical tool
  to implement \textit{ab initio} calculations
  of ground-state energies and densities in many-electron systems.
  There have been attempts to extend the KS theory
  so that it could describe other physical quantities,
  or it could be applied to other many-fermion systems.
  By generalizing and reformulating the KS theory
  in terms of the 1-body density matrix,
  we expose the minimal composition of the theory
  that enables the reduction of the many-fermion problem to the s.p. equations.
  Based on the reformulation, several basic issues are reconsidered.
  The $v$- and $N$-representabilities for the KS theory
  are distinguished from those for the Hohenberg-Kohn theorem.
  Criteria for the extendability of the KS theory are addressed.
\end{abstract}


\maketitle



\section{Introduction\label{sec:intro}}

The density-functional theory (DFT) successfully described
many-electron systems ranging from atoms and molecules
to solids~\cite{Parr-Yang_1989,Giuliani-Vignale_2005,Engel-Dreizler_2011,
  Sahni_2016},
in which the influence of the ions is handled as external fields
via the Born-Oppenheimer approximation.
Located at its center,
the Hohenberg-Kohn (HK) theorem~\cite{Hohenberg-Kohn}
has provided the firm basis of the DFT,
and the Kohn-Sham (KS) theory~\cite{Kohn-Sham}
has made the DFT practical for electronic systems.
They have supplied an \textit{ab initio} theory
for ground-state (g.s.) energy $E_0$.
The energy of system $E$ is represented as a functional
of the local density $n(\mathbf{r})$,
and $E_0$ is obtained via the variation of the energy density functional (EDF).

The KS theory has been attempted to be extended
in various directions.
In electronic systems, the EDFs including variables
additional to $n(\mathbf{r})$ were developed:
for instance, the spin DFT~\cite{vonBarth-Hedin_1972,Gunnarsson-Lundqvist_1976,
  Jacob-Reiher_2012},
the current DFT~\cite{Vignale-Rasolt_1987,Vignale-Rasolt_1988},
the DFT for superconductors~\cite{Oliveira-Gross-Kohn_1988,Lueders-etal_2005},
and the density-matrix functional theory~\cite{Donnely-Parr_1978,
  Zumbach-Maschke_1985,Schindlmayr-Godby_1995,LopezSandoval-Pastor_2000,
  Requist-Pankratov_2008,Pernal-Giesbertz_2016}.
A systematic method to involve additional quantities
has been proposed~\cite{Higuchi-Higuchi_2004}.
The DFT has also been extended
to finite-temperature problems~\cite{Mermin_1965}
and to excitations via a time-dependent formalism~\cite{Runge-Gross_1984,
  Gross-Kohn_1985},
although we focus on the g.s. properties in this paper.
Another direction is to extend the KS theory to other many-fermion systems.
For atomic nuclei composed of nucleons,
the self-consistent mean-field (SCMF) calculations
have been implemented~\cite{Vautherin-Brink}
by using effective interactions in which correlation effects are incorporated.
The most popular form of effective interactions
is the Skyrme interaction~\cite{Skyrme_1959},
representing energy via quasi-local currents,
\textit{i.e.}, currents that depend only on a position $\mathbf{r}$,
by attributing influences of non-locality to momentum-dependence.
These SCMF approaches have been reinterpreted
in the context of the DFT~\cite{Petkov-Stoitsov_1991,Colo_2020}.
Calculations using finite-range effective interactions,
which lead to non-local EDFs,
have been developed as well~\cite{Gogny_D1,Nakada_2003}.
The SCMF approaches incorporating relativity~\cite{Serot-Walecka_1986}
are also renamed the covariant DFT~\cite{Ring_2016}.

The KS theory's remarkable and essential property
is that the many-fermion problem is reduced to
a set of single-particle (s.p.) equations.
In this paper, we shall generalize and reformulate the KS theory,
aiming to expose the minimal composition of the theory,
\textit{i.e.}, with what ingredients at minimum
the many-fermion problem can be solved by s.p. equations.
Through the reformulation,
criteria for the extendability of the theory will be elucidated.
They help to judge what physical quantities can be covered by extending it
and to what system it can be extended.

The so-called orbital-free (OF) DFT,
which has a root in the historical work
by Thomas and Fermi~\cite{Thomas_1927,Fermi_1928},
has been developed~\cite{Deb-Chattaraj_1988,Deb-Chattaraj_1992,
  Pearson-Smargiassi-Madden_1993}
to reduce the computational cost further.
Whereas the OF DFT is often considered
an approximation of the KS DFT,
it does not need s.p. equations and should be distinguished from the KS theory
in this respect.
As one of our central issues is the emergence of the s.p. equations,
we will not handle the OF DFT in this paper, except for a brief referral.

We restrict our discussion to systems where the particle number $N$ is conserved.
An arbitrary normalized state in the system is represented by $|\Psi\rangle$.
Denoting the Hamiltonian of this system by $H$,
the energy expectation value is regarded
as a functional of $|\Psi\rangle$,
\begin{equation}
  E[\Psi] := \langle\Psi|H|\Psi\rangle\,.
  \label{eq:EDF0}
\end{equation}
The variational principle tells us
that $E_0$ is equal to the minimum\footnote{
  The expression $\min f(x)$ means the global minimum of $f(x)$ as usual,
  throughout this paper.}
of $E[\Psi]$ with respect to $|\Psi\rangle$,
\begin{equation}
  E_0 = \min_{\Psi\to N} E[\Psi]\,,
  \label{eq:min_E}
\end{equation}
where the subscript $\Psi\to N$ stands for the condition
that the state $|\Psi\rangle$ should belong to the $N$-particle system.
However, the underlying Hamiltonian is not well established
in certain systems;
the atomic nuclei are among them.
For this reason, $E[\Psi]$ and the variational principle \eqref{eq:min_E}
are placed at the starting point in the arguments below,
not necessarily presuming $H$.
Moreover, we shall present the formulation
in terms of the 1-body density matrix,
which carries properties of $|\Psi\rangle$.
While $n(\mathbf{r})$ is adopted as the principal variable for the variation
in the original HK theorem,
we merely assume that the principal variables
are all s.p. quantities,
viz., functions of the 1-body density matrix,
not restricting to $n(\mathbf{r})$.

\section{Hartree-Fock approximation\label{sec:SCMF}}

The KS theory significantly overlaps with the SCMF approximations.
Both approaches reduce many-fermion problems
to a set of s.p. equations.
In this section we review the Hartree-Fock (HF) approximation,
which is a typical and rigorous example of the SCMF theories,
in terms of the 1-body density matrix~\cite{Ring-Schuck}.
The arguments here will clarify the basic structure of the SCMF theories,
which is possessed by the KS theory as well.
Another example, the Hartree-Fock-Bogolyubov (HFB) approximation,
is summarized in Appendix~\ref{app:HFB}.


Let $\{\phi_k\}$ represent a complete orthonormal set of s.p. bases.
The creation and annihilation operators of the basis state $k$
are denoted by $a^\dagger_k$ and $a_k$.
The 1-body density matrix\footnote{
  This density matrix is called first-order reduced density matrix
  in some literature.}
$\varrho_{k\ell}$
for an arbitrary normalized state $|\Psi\rangle$ is defined by
\begin{equation}
  \varrho_{k\ell}=\varrho_{k\ell}[\Psi]
  :=\langle\Psi|a^\dagger_\ell a_k|\Psi\rangle
  =\varrho_{\ell k}^\ast\,.
\end{equation}
The $\nu$-body density matrix
$\langle\Psi|a^\dagger_{\ell_1}a^\dagger_{\ell_2}\cdots a^\dagger_{\ell_\nu}
a_{k_\nu}\cdots a_{k_2}a_{k_1}|\Psi\rangle$
is decomposed into products of $\varrho_{k\ell}$
and the correlation functions $\mathcal{C}^{(\nu')}$ ($\nu'=2,3,\cdots,\nu$),
which obey the Bogolyubov-Born-Green-Kirkwood-Yvon hierarchy~\cite{
  Reichl_1980}.
For instance, the 2-body correlation function is defined by
\begin{equation}
  \mathcal{C}^{(2)}_{kk'\ell\ell'}
  := \langle\Psi|a^\dagger_\ell a^\dagger_{\ell'} a_{k'} a_k|\Psi\rangle
  - \varrho_{k\ell} \varrho_{k'\ell'}
  + \varrho_{k\ell'} \varrho_{\ell k'}\,.
\end{equation}
$E[\Psi]$ can be represented by $\varrho$ and $\mathcal{C}^{(\nu)}$,
\begin{equation}
  E[\Psi]
  = E[\varrho,\mathcal{C}^{(2)},\mathcal{C}^{(3)},\cdots,\mathcal{C}^{(N)}]\,.
\end{equation}

By neglecting $\mathcal{C}^{(\nu)}$ ($\nu\geq 2$),
the energy functional in the HF approximation is obtained,
$E[\Psi]\approx E^\mathrm{HF}[\varrho]$.
Variation of $E^\mathrm{HF}[\varrho]$ for $\varrho_{k\ell}$
yields a s.p. Hamiltonian (HF Hamiltonian),
\begin{equation}
  h^\mathrm{HF}_{k\ell}
  := \frac{\partial E^\mathrm{HF}}{\partial\varrho_{\ell k}}
  = \big(h^\mathrm{HF}_{\ell k}\big)^\ast\,.
  \label{eq:h_HF}
\end{equation}
The diagonalization of $h^\mathrm{HF}$
with an appropriate unitary transformation $\mathcal{U}^\mathrm{HF}$
yields
\begin{equation}
  \sum_\ell h^\mathrm{HF}_{k\ell}\,\mathcal{U}^\mathrm{HF}_{\ell i}
  = \epsilon^\mathrm{HF}_i\,\mathcal{U}^\mathrm{HF}_{ki}\,,
  \label{eq:HF-eq}
\end{equation}
which is identical to the HF equation,
determining s.p. energy $\epsilon^\mathrm{HF}_i$
and wave-function $\varphi^\mathrm{HF}_i
=\sum_k \mathcal{U}^\mathrm{HF}_{ki}\,\phi_k$.

The HF state $|\Phi^\mathrm{HF}\rangle$ is obtained
by minimizing $E^\mathrm{HF}$ with respect to $\varrho_{k\ell}$
under the particle-number condition,
\begin{equation}
  \mathrm{tr}(\varrho) = N\,.
  \label{eq:N-cond}
\end{equation}
The trace in Eq.~\eqref{eq:N-cond} is taken in the s.p. space.
To implement the minimization,
Eq.~\eqref{eq:N-cond} is handled with the Lagrange multiplier $\mu$,
modifying $E^\mathrm{HF}$ as
\begin{equation}
  \tilde{E}^\mathrm{HF}
  := E^\mathrm{HF} - \mu \big[\mathrm{tr}(\varrho) - N\big]\,.
  \label{eq:Emod_HF}
\end{equation}
Variation of $\tilde{E}^\mathrm{HF}$ yields
\begin{equation}\begin{split}
  \delta\tilde{E}^\mathrm{HF}
  &= \sum_{k\ell} \frac{\partial E^\mathrm{HF}}{\partial\varrho_{\ell k}}\,
  \delta\varrho_{\ell k} - \mu \sum_k \delta\varrho_{kk}
  - \delta\mu \big[\mathrm{tr}(\varrho) - N\big] \\
  &= \mathrm{tr}\big[(h^\mathrm{HF}-\mu)\,\delta\varrho\big]
  - \delta\mu \big[\mathrm{tr}(\varrho) - N\big]\,.
\end{split}\label{eq:var_E_HF}
\end{equation}
We should have\footnote{
  Note that $\delta\tilde{E}^\mathrm{HF}=0$ is not necessarily satisfied
  because $\varrho_{k\ell}$ does not vary freely,
  restricted by the bounds of Eq.~\eqref{eq:Pauli}.
  See also Footnote~\protect\ref{foot:var_eq}.}
$\delta\tilde{E}^\mathrm{HF}\geq 0$ at the minimum of $E^\mathrm{HF}$.

The Pauli principle requires
\begin{equation}
  0\leq\varrho_{ii}\leq 1\quad\mbox{for any representation}\,.
  \label{eq:Pauli}
\end{equation}
By combining $\delta\tilde{E}^\mathrm{HF}\geq 0$ with Eq.~\eqref{eq:Pauli},
it is concluded that $|\Phi^\mathrm{HF}\rangle$
should be a state corresponding to a single Slater determinant,
as proven below.
In $\delta\tilde{E}^\mathrm{HF}$ of Eq.~\eqref{eq:var_E_HF},
the $\delta\mu$ term vanishes as long as Eq.~\eqref{eq:N-cond} is fulfilled.
Suppose that $\{\varrho^\mathrm{HF}_{k\ell}
\,(:=\varrho_{k\ell}[\Phi^\mathrm{HF}])\}$
gives the minimum of $E^\mathrm{HF}$ under the constraint \eqref{eq:N-cond},
which is equal to the minimum of $\tilde{E}^\mathrm{HF}$.
The s.p. energy $\epsilon^\mathrm{HF}_i$
and wave-function $\varphi^\mathrm{HF}_i$ are determined accordingly.
It is convenient to work with the s.p. state $i$
instead of the basis $k$ in the vicinity of the HF solution.
In this vicinity, Eq.~\eqref{eq:var_E_HF} is expressed as
\begin{equation}
  \delta \tilde{E}^\mathrm{HF}
  = \sum_i (\epsilon^\mathrm{HF}_i-\mu)\,\delta\varrho_{ii}\,.
  \label{eq:var_E_HF-2}
\end{equation}
The requirement $\delta \tilde{E}^\mathrm{HF}\geq 0$
at the minimum of $E^\mathrm{HF}$ derives
\begin{equation}
  \left\{\begin{array}{ll}
  \delta\varrho_{ii}\geq 0\quad
  & \mbox{for $i$ with }\epsilon^\mathrm{HF}_i>\mu\\
  \delta\varrho_{ii}\leq 0 & \mbox{for $i$ with }\epsilon^\mathrm{HF}_i<\mu
  \end{array}\right.\,.
  \label{eq:var_1bd_HF}
\end{equation}
Combined with \eqref{eq:Pauli},
Eq.~\eqref{eq:var_1bd_HF} leads to
\begin{equation}
  \left\{\begin{array}{ll}
  \varrho^\mathrm{HF}_{ii}=0\quad
  & \mbox{for $i$ with }\epsilon^\mathrm{HF}_i>\mu\\
  \varrho^\mathrm{HF}_{ii}=1 & \mbox{for $i$ with }\epsilon^\mathrm{HF}_i<\mu
  \end{array}\right.\,,
  \label{eq:1bd_HF}
\end{equation}
so that negative $\delta\varrho_{ii}$ should not be allowed
for $\epsilon^\mathrm{HF}_i>\mu$
and likewise for $\epsilon^\mathrm{HF}_i<\mu$.
Since $\varrho_{ii}$ is the occupation number on the s.p. state $i$,
the lowest $N$ s.p. states are occupied in $|\Phi^\mathrm{HF}\rangle$
unless there exist s.p. levels with $\epsilon^\mathrm{HF}_i=\mu$,
which could be degenerate.
Even if there are several levels having $\epsilon^\mathrm{HF}_i=\mu$,
the occupation number on them can be set integer,
viz., $0$ or $1$,
through a unitary transformation among the degenerate s.p. states.
Therefore $|\Phi^\mathrm{HF}\rangle$
can be represented by a single Slater determinant,
which is formed by $\{\varphi^\mathrm{HF}_i\}$ with $\varrho^\mathrm{HF}_{ii}=1$.
As the $\epsilon^\mathrm{HF}_i=\mu$ levels do not influence essential points
but make arguments lengthy,
we shall assume $\epsilon^\mathrm{HF}_i\ne\mu$ for any $i$ below. 

Because of Condition~\eqref{eq:Pauli},
$\varrho^\mathrm{HF}_{ii}$ of Eq.~\eqref{eq:1bd_HF} has reached
either its maximum or minimum
that can be obtained by any unitary transformation,
implying that $\varrho^\mathrm{HF}$ is diagonalized in this representation.
When its eigenvalues are only $0$ and $1$,
the density matrix $\varrho$ is characterized by the idempotency\footnote{
  By regarding Eq.~\eqref{eq:idempotency} as a constraint
  and defining $\tilde{\tilde{E}}^\mathrm{HF}
  := \tilde{E}^\mathrm{HF} - \sum_{k\ell}\lambda_{k\ell}\,
  (\varrho^2-\varrho)_{k\ell}$
    with the Lagrange multiplier $\lambda_{k\ell}$,
  the minimum of $E^\mathrm{HF}$ can be searched
  via $\delta\tilde{\tilde{E}}^\mathrm{HF}=0$,
  compatible with the HF equation.
  \label{foot:var_eq}},
\begin{equation}
  \varrho^2 = \varrho\,,\quad\mbox{viz.,}\quad
  \sum_m \varrho_{km}\,\varrho_{m\ell} = \varrho_{k\ell}\,.
  \label{eq:idempotency}
\end{equation}
A state given by a single Slater determinant $|\Phi\rangle$ provides
$\varrho_{k\ell}=\langle\Phi|a^\dagger_\ell a_k|\Phi\rangle$
that fulfills Eq.~\eqref{eq:idempotency},
as is evident by representing it through the s.p. states
$\{\varphi_i;i=1,2,\cdots,N\}$ forming $|\Phi\rangle$.
Conversely, if $\{\varrho_{k\ell}\}$ satisfies Eq.~\eqref{eq:idempotency},
the corresponding state is given by a Slater determinant.
This assertion is easily proven by constructing the Slater determinant:
take the s.p. states $\{\varphi_i\}$
that diagonalize $\varrho$
and pick up the states corresponding to the eigenvalue $1$.
Thus, $\varrho^\mathrm{HF}$ satisfying Eq.~\eqref{eq:idempotency}
and $|\Phi^\mathrm{HF}\rangle$ is represented by a single Slater determinant.
Let us denote the full Hilbert space of the $N$-particle system
by $\mathcal{V}_\mathrm{full}$,
and the subspace composed of the states of the Slater determinants
by $\mathcal{V}_\mathrm{idem}$,
which are characterized by Eq.~\eqref{eq:idempotency}.
It is noted that,
although $E^\mathrm{HF}[\varrho]$ is defined in $\mathcal{V}_\mathrm{full}$,
the wave-function necessarily falls on $\mathcal{V}_\mathrm{idem}$,
as the result of the variation.
This fact is equivalent to the conclusion in Ref.~\cite{Lieb_1981}.

\section{Hohenberg-Kohn theorem\label{sec:HK}}

We present the Hohenberg-Kohn (HK) theorem in this section.
Whereas it is no more than a summary of many previous works,
we state and discuss the HK theorem
in a generalized context,
preparing to reformulate the KS theory.

Consider a set of variables $Q^A$,
which are called principal variables in this article.
The superscript $A$ distinguishes the variables,
which can be continuous.
For brevity, the set of $Q^A$ is denoted by $\mathbf{Q}$.
In the original HK theorem~\cite{Hohenberg-Kohn},
$\mathbf{Q}$ is the local density $n(\mathbf{r})
:=\langle\Psi|\sum_{i=1}^N \delta(\mathbf{r}-\mathbf{r}_i)|\Psi\rangle$,
where $i$ is the index of particles\footnote{
  The index $i$ is commonly used for labeling particles
  and s.p. states after diagonalization,
  which correspond to each other.
  }.
As mentioned in Introduction,
additional variables have been introduced in $\mathbf{Q}$
when extending the DFT.
In Ref.~\cite{Gilbert},
the HK theorem has extensively been proven for the whole 1-body density matrix,
$\mathbf{Q}=\{\varrho_{k\ell}\}$.

The HK theorem (the so-called 2nd theorem)
asserts the existence of the functional $E^\mathrm{HK}[\mathbf{Q}]$,
whose minimum gives exact g.s. energy and corresponding $\mathbf{Q}$.
The theorem was initially proven
by showing the one-to-one correspondence
of $\mathbf{Q}=n(\mathbf{r})$ and the external potential (the 1st HK theorem),
which is equivalent to the Legendre transformation.
The Legendre transformation was also applied
in the effective action formalism,
deriving the HK theorem in an elegant manner~\cite{Fukuda-etal_1994,
  Valiev-Fernando_1997}.
However, we follow Levy's idea of the constrained search~\cite{Levy},
which derives a universal energy functional of $\mathbf{Q}$
and provides a basis for the KS theory.
The HK energy functional can be defined by
\begin{equation}
  E^\mathrm{HK}[\mathbf{Q}] := \min_{\Psi\to\mathbf{Q}} E[\Psi]\,.
  \label{eq:EHK}
\end{equation}
Here the expression $\Psi\to\mathbf{Q}$ indicates
the constraint that $|\Psi\rangle$ gives a fixed $\mathbf{Q}$,
under which the minimization with respect to $|\Psi\rangle$ is carried out.
The energy functional of $\mathbf{Q}$ is extensively called
EDF in this paper,
though $\mathbf{Q}$ is not necessarily the density.
Equation~\eqref{eq:EHK} guarantees the uniqueness of the EDF
except for degeneracy,
as it is defined as the global minimum under $\Psi\to\mathbf{Q}$.
For later convenience,
the state $|\Psi\rangle$ that minimizes the energy for a fixed $\mathbf{Q}$
is denoted by $|\Psi^\mathrm{HK}_\mathbf{Q}\rangle$;
\begin{equation}
  E^\mathrm{HK}[\mathbf{Q}]\,\big(= \langle\Psi^\mathrm{HK}_\mathbf{Q}|
  H|\Psi^\mathrm{HK}_\mathbf{Q}\rangle\big)
  = E[\Psi^\mathrm{HK}_\mathbf{Q}]\,.
  \label{eq:EHK2}
\end{equation}
The minimum of $E^\mathrm{HK}[\mathbf{Q}]$ with respect to $\mathbf{Q}$
in the $N$-particle space
gives the exact g.s. energy,
\begin{equation}
  E_0 = \min_{\mathbf{Q}\to N} E^\mathrm{HK}[\mathbf{Q}]\,.
  \label{eq:min-EHK}
\end{equation}
The values of $\mathbf{Q}$ that give the minimum are also exact.
The HK theorem means that the minimization of \eqref{eq:min_E}
can be separated into two steps, \eqref{eq:EHK} and \eqref{eq:min-EHK}.

The representability problem should not be discarded,
although this issue is not easy to be explored thoroughly.
It is not trivial whether $\mathcal{V}_\mathrm{full}$,
the full Hilbert space spanned by $|\Psi\rangle$ with $N$ particles,
covers the whole domain of $\mathbf{Q}$
(viz., the $N$-representability)~\cite{Coleman}.
This problem may limit the choice of $\mathbf{Q}$.
If $\mathbf{Q}=n(\mathbf{r})$,
it has been shown~\cite{Harriman}
that there exists $|\Psi\rangle$ for any $n(\mathbf{r})$
that fulfills $n(\mathbf{r})\geq 0$ and $\int d^3r\,n(\mathbf{r}) = N$.

In the 1st HK theorem,
$n(\mathbf{r})$ is linked to the external s.p. potential
via the Legendre transformation.
In this case, the $v$-representability,
\textit{i.e.}, the existence of the s.p. potential conjugate
to any physical $n(\mathbf{r})$, should be assured.
It has been pointed out that $v$-representability
is not always satisfied~\cite{Levy_1982}.
Furthermore, the EDF via the Legendre transformation
yields the global minimum only in a restricted region of $\mathbf{Q}$.
Examples in the spin DFT are found in Refs.~\cite{vonBarth-Hedin_1972,
  Capelle-Vignale_2001},
and presented in Sec.~\ref{subsec:electrons}.
Thus the Legendre transformation derives an EDF
that yields the g.s. energy only in a limited domain of $\mathbf{Q}$.
Notice that the EDF obtained from the Legendre transformation
is not equivalent to $E^\mathrm{HK}[\mathbf{Q}]$ of Eq.~\eqref{eq:EHK}
everywhere.
They are different in the domain
where the EDF of the Legendre transformation does not give the global minimum.

While the constrained search circumvents this problem,
it induces another problem concerning differentiability.
To carry out the minimization of Eq.~\eqref{eq:min-EHK},
the condition $\delta E^\mathrm{HK}[\mathbf{Q}]\geq 0$
(or $\delta E^\mathrm{HK}[\mathbf{Q}]=0$) is imposed,
where the expression $\delta$ contains
differentiation with respect to $\mathbf{Q}$,
though we postpone specifying it concretely.
Therefore $E^\mathrm{HK}[\mathbf{Q}]$ should be differentiable
for this practical purpose.
As will be shown in Sec.~\ref{sec:KS},
a s.p. potential emerges by the differentiation.
Indeed, the functional derivative of the EDF for $n(\mathbf{r})$
yields a s.p. potential
in the original discussion on the HK theorem.
In this respect,
the $v$-representability remains after the constrained search,
transferred to the differentiability of $E^\mathrm{HK}[\mathbf{Q}]$.
The importance of differentiability, or regularity,
has been pointed out in Refs.~\cite{Capelle-Vignale_2001,Kvaal-etal_2014}.
As illustrated in Appendix~\ref{app:2-step},
regularity is often lost if optimization is separated into two steps.
The two-step optimization is essential for the universality
of $E^\mathrm{HK}[\mathbf{Q}]$
but damages regularity as a function (or functional) of $Q^A$ in most cases.
Without this regularity,
the procedure of Eq.~\eqref{eq:min-EHK} is impractical.

\section{Kohn-Sham theory\label{sec:KS}}

The Kohn-Sham (KS) theory is now reformulated and generalized\footnote{
  Although it is a generalization of the original KS theory,
  we keep calling it \textit{KS theory},
  as some existing extensions may implicitly be subject
  to this generalization.}.
The irregularity of $E^\mathrm{HK}[\mathbf{Q}]$ mentioned above
may occur as a quantum effect,
as in the shell structure of atoms and nuclei.
The argument arriving at Eq.~\eqref{eq:1bd_HF} is indicative
of this problem.
Even if the particle number $N$ were a continuous variable,
not restricted to integers,
the condition \eqref{eq:Pauli} should lead to an irregularity
because particles start occupying another s.p. state
after one of the $\varrho_{ii}$'s reaches unity.
This property originating from the Pauli principle
is the source of the shell effect.
Influence of the shell structure on the regularity of the EDF
is discussed in Appendix~\ref{app:HO-EDF}.
This perception concerning the shell effect
motivates us to deal with $\varrho$ explicitly,
so that the EDF could come well-behaved.

If $N$ is very large and the quantum effects are not strong,
the irregularity due to the shell structure may be negligible.
It is then possible to directly apply a variational calculation
to $E^\mathrm{HK}[\mathbf{Q}]$,
deriving the orbital-free DFT,
although it is not the current interest.

\subsection{Kohn-Sham equation\label{subsec:KS-eq}}

In the KS theory~\cite{Kohn-Sham},
which realizes the HK theorem,
the energy variation is reduced to equations
defining a set of s.p. orbitals, called KS orbitals.
This is a striking feature of the KS theory.
To disclose what ingredients at minimum
enable to solve the many-fermion problem by s.p. equations,
the equations are rederived
by means of the 1-body density matrix $\varrho$.

Impose $Q^A=Q^A[\varrho]$,
\textit{i.e.}, all $Q^A$'s depend solely on the 1-body density matrix $\varrho$,
not on $\mathcal{C}^{(\nu)}$ ($\nu\geq 2$).
Moreover, $Q^A$'s have to be independent of one another
as a function of $\varrho$.
The energy $E[\Psi]
=E[\varrho,\mathcal{C}^{(2)},\mathcal{C}^{(3)},\cdots,\mathcal{C}^{(N)}]$
is partitioned into two parts,
regular and irregular parts with respect to $\mathbf{Q}$.
The irregular part is assumed to be represented
only with $\varrho$, without $\mathcal{C}^{(\nu)}$ ($\nu\geq 2$),
and to be differentiable with respect to $\varrho$
though not with respect to $\mathbf{Q}$.
The irregular part may explicitly contain $Q^A$ without contradiction,
because $Q^A=Q^A[\varrho]$.
The partitioning is then expressed as
\begin{equation}
  E[\Psi] = E^\mathrm{KS}_\mathrm{irr}\big[\varrho;\mathbf{Q}[\varrho]\big]
  + E_\mathrm{reg}[\Psi]\,.
  \label{eq:EHK-sep0}
\end{equation}
This partitioning may correspond to that of the underlying Hamiltonian,
\begin{equation}
  H = H_\mathrm{irr} + H_\mathrm{reg}\,;\quad
  E^\mathrm{KS}_\mathrm{irr}\big[\varrho;\mathbf{Q}[\varrho]\big]
  = \langle\Psi|H_\mathrm{irr}|\Psi\rangle\,,~
  E_\mathrm{reg}[\Psi] = \langle\Psi|H_\mathrm{reg}|\Psi\rangle\,,
  \label{eq:Hamil-sep0}
\end{equation}
although it is not requisite.

The individual theory is not defined
until fixing $\mathbf{Q}$ and the partitioning
(viz., $E^\mathrm{KS}_\mathrm{irr}$).
In the standard KS equation, $\mathbf{Q}=n(\mathbf{r})$
and $E^\mathrm{KS}_\mathrm{irr}$ corresponds to the kinetic energy term
so that the shell effects could be tractable.
This also holds in the example of Appendix~\ref{app:HO-EDF}.
However, the source of the irregularity
is not necessarily restricted to the kinetic energy in general cases.
The choice of $\mathbf{Q}$ and $E^\mathrm{KS}_\mathrm{irr}$ is not always obvious;
it may count on physical insight.
Still, the structure of the KS theory can be discussed
without specifying $\mathbf{Q}$ and $E^\mathrm{KS}_\mathrm{irr}$,
as long as the influence of $\mathcal{C}^{(\nu)}$ ($\nu\geq 2$) on them
is masked.

With the partitioning of Eq.~\eqref{eq:EHK-sep0},
the result of the constrained search of Eq.~\eqref{eq:EHK} is written as
\begin{equation}
  E^\mathrm{HK}[\mathbf{Q}]
  = E^\mathrm{KS}_\mathrm{irr}\big[\varrho[\Psi^\mathrm{HK}_\mathbf{Q}];
    \mathbf{Q}\big]
  + E^\mathrm{HK}_\mathrm{reg}[\mathbf{Q}]\,;\quad
  E^\mathrm{HK}_\mathrm{reg}\big[\mathbf{Q}]
  := E_\mathrm{reg}\big[\Psi^\mathrm{HK}_\mathbf{Q}\big]\,.
  \label{eq:EHK-sep1}
\end{equation}
Whereas $E^\mathrm{KS}_\mathrm{irr}$ in Eq.~\eqref{eq:EHK-sep1}
is not determined until knowing $\varrho[\Psi^\mathrm{HK}_\mathbf{Q}]$,
it is a difficult task to obtain $\varrho[\Psi^\mathrm{HK}_\mathbf{Q}]$,
not more accessible than $|\Psi^\mathrm{HK}_\mathbf{Q}\rangle$.
A practical way is to minimize $E^\mathrm{KS}_\mathrm{irr}$ for $\varrho$
under the constraint that $\varrho$ should provide a given $\mathbf{Q}$,
$\min_{\varrho\to\mathbf{Q}} E^\mathrm{KS}_\mathrm{irr}[\varrho;\mathbf{Q}]$.
The resultant $\varrho$ is not equal
to $\varrho[\Psi^\mathrm{HK}_\mathbf{Q}]$,
and thereby
$\min_{\varrho\to\mathbf{Q}} E^\mathrm{KS}_\mathrm{irr}[\varrho;\mathbf{Q}]
\ne E^\mathrm{KS}_\mathrm{irr}\big[\varrho[\Psi^\mathrm{HK}_\mathbf{Q}];
  \mathbf{Q}\big]$.
However, this deviation is assumed to be a regular function of $\mathbf{Q}$.
Then, the deviation can be incorporated into the regular part of the EDF,
deriving the modified partitioning as
\begin{equation}\begin{split}
  & E^\mathrm{HK}[\mathbf{Q}]
  = \min_{\varrho\to\mathbf{Q}} E^\mathrm{KS}_\mathrm{irr}[\varrho;\mathbf{Q}]
  + E^\mathrm{KS}_\mathrm{reg}[\mathbf{Q}]\,;\\
  &\quad E^\mathrm{KS}_\mathrm{reg}[\mathbf{Q}]
  := E^\mathrm{HK}_\mathrm{reg}[\mathbf{Q}]
  + \Big(E^\mathrm{KS}_\mathrm{irr}\big[\varrho[\Psi^\mathrm{HK}_\mathbf{Q}];
    \mathbf{Q}\big]
  - \min_{\varrho\to\mathbf{Q}} E^\mathrm{KS}_\mathrm{irr}[\varrho;\mathbf{Q}]
  \Big)\,.
\end{split}\label{eq:EHK-sep2}
\end{equation}

Although the principal variables $\mathbf{Q}$
should be a function of the density matrix $\varrho$,
it is convenient to handle them
as if they were independent in $E^\mathrm{KS}_\mathrm{irr}$.
To do it,
$Q^A$ is replaced by an auxiliary variable $q^A$,
with setting $q^A=Q^A$ as a constraint.
The EDF within the KS scheme is now defined as follows,
with the Lagrange multiplier $\Lambda^\mathrm{KS}_A$,
\begin{equation}
  E^\mathrm{KS}[\varrho;\mathbf{q}] =
  E^\mathrm{KS}_\mathrm{irr}[\varrho;\mathbf{q}]
  + E^\mathrm{KS}_\mathrm{reg}[\mathbf{q}]
  - \sum_A \Lambda^\mathrm{KS}_A\,\Big(q^A - Q^A[\varrho]\Big)\,.
  \label{eq:KS-EDF}
\end{equation}
This procedure is analogous to the Legendre transformation
used in Refs.~\cite{Fukuda-etal_1994,Valiev-Fernando_1997}.
Then, after the treatment analogous to Eq.~\eqref{eq:Emod_HF},
\begin{equation}
  \tilde{E}^\mathrm{KS}
  := E^\mathrm{KS} - \mu \big[\mathrm{tr}(\varrho) - N\big]\,,
  \label{eq:Emod_KS}
\end{equation}
the condition $\delta\tilde{E}^\mathrm{KS}[\varrho;\mathbf{q}]\geq 0$
should be fulfilled at the energy minimum,
where the variation is performed for $\varrho$, $\mathbf{q}$ and $\mu$.
The variation of $E^\mathrm{KS}$ for $q^A$ yields
\begin{equation}
  \Lambda^\mathrm{KS}_A
  = \frac{\partial E^\mathrm{KS}_\mathrm{irr}[\varrho;\mathbf{q}]}{\partial q^A}
  + \frac{\partial E^\mathrm{KS}_\mathrm{reg}[\mathbf{q}]}{\partial q^A}\,.
  \label{eq:KSpot}
\end{equation}
Here the differentiation with respect to $q^A$ is expressed
by the partial derivative,
but it should read as a functional derivative
when $A$ is continuous. 
The variation provides s.p. Hamiltonian,
\begin{equation}\begin{split}
  h^\mathrm{KS}_{k\ell}
    &:= \frac{\partial E^\mathrm{KS}}{\partial\varrho_{k\ell}}
  = \frac{\partial E^\mathrm{KS}_\mathrm{irr}[\varrho;\mathbf{q}]}
       {\partial\varrho_{k\ell}}
       + \Gamma^\mathrm{KS}_{k\ell}\,;\\
  \Gamma^\mathrm{KS}_{k\ell}
  &:= \sum_A \Lambda^\mathrm{KS}_A\,\frac{\partial Q^A}{\partial\varrho_{k\ell}}
  = \sum_A \bigg(\frac{\partial E^\mathrm{KS}_\mathrm{irr}[\varrho;\mathbf{q}]}
  {\partial q^A}
  + \frac{\partial E^\mathrm{KS}_\mathrm{reg}[\mathbf{q}]}{\partial q^A}\bigg)\,
  \frac{\partial Q^A}{\partial\varrho_{k\ell}}\,.
\end{split}\label{eq:KS-hamil}\end{equation}
$\Gamma^\mathrm{KS}$ is the KS potential.
It is reasonable to assume that $(Q^{A})^\ast$ is also contained in $\mathbf{Q}$
for a complex $Q^A$.
Then $h^\mathrm{KS}$ is hermitian and therefore diagonalizable,
having real eigenvalues,
\begin{equation}
  \sum_\ell h^\mathrm{KS}_{k\ell}\,\mathcal{U}^\mathrm{KS}_{\ell i}
  = \epsilon^\mathrm{KS}_i\,\mathcal{U}^\mathrm{KS}_{ki}\,,
  \label{eq:KS-eq}
\end{equation}
similarly to the HF equation \eqref{eq:HF-eq}.
This is the KS equation,
and the s.p. state diagonalizing $h^\mathrm{KS}$,
$\varphi^\mathrm{KS}_i=\sum_k \mathcal{U}^\mathrm{KS}_{ki}\,\phi_k$,
is the KS orbital.
Because of the identical mathematical structure,
all the arguments on the solutions of the HF equation in Sec.~\ref{sec:SCMF}
hold for those of the KS equation \eqref{eq:KS-eq} as they are.
The resultant 1-body density matrix $\varrho^\mathrm{KS}$
satisfies Eq.~\eqref{eq:idempotency},
corresponding to the state $|\Phi^\mathrm{KS}\rangle$
in which the lowest $N$ KS orbitals are occupied.
However, while all $Q^A$'s are physical,
individual $\varrho^\mathrm{KS}_{k\ell}$
(equivalently, $\mathcal{U}^\mathrm{KS}_{ki}$ and $\varphi^\mathrm{KS}_i$)
is artificial, losing clear physical meaning,
as is recognized from the derivation.

In the conventional formulation of the KS theory,
the subspace $\mathcal{V}_\mathrm{idem}$ is considered from the beginning
by introducing the non-interacting reference system and the KS orbitals.
On the other hand,
the functional $E^\mathrm{KS}[\varrho;\mathbf{q}]$ is here defined
in the full Hilbert space $\mathcal{V}_\mathrm{full}$.
Hence it can be identified as an effective Hamiltonian,
although it depends on $\mathbf{q}$ and is not necessarily representable
in the second-quantized form.
Solutions of the KS equation fall on $\mathcal{V}_\mathrm{idem}$ as a result.
Still, it is legitimate to state
that the idempotent $\varrho$ ($\varrho^\mathrm{KS}$, to be precise)
is implicitly assumed
when postulating that $E^\mathrm{KS}_\mathrm{irr}$ does not depend
on $\mathcal{C}^{(\nu)}$ ($\nu\geq 2$).

As already mentioned, the HK theorem was extended
to the case $\mathbf{Q}=\{\varrho_{k\ell}\}$~\cite{Gilbert},
above which the density-matrix functional theory (DMFT) was exploited~\cite{
  Donnely-Parr_1978,Zumbach-Maschke_1985,Schindlmayr-Godby_1995,
  LopezSandoval-Pastor_2000,Requist-Pankratov_2008,Pernal-Giesbertz_2016}.
The present reformulation should be distinguished from the DMFT.
It is noticed that $E^\mathrm{HK}[\varrho_{k\ell}]$
(\textit{i.e.}, $\mathbf{Q}=\{\varrho_{k\ell}\}$) itself
should not be handled in the KS theory
because the 1-body density matrix of the KS theory
$\varrho\,(=\varrho^\mathrm{KS})$
cannot be physical,
as will be discussed in Sec.~\ref{sec:EDF}.
The ensemble representability~\cite{Valone_1980} is irrelevant
to $\varrho$ here.

If $E^\mathrm{KS}_\mathrm{irr}[\varrho;\mathbf{q}]$ consists of two parts,
one which linearly depends on $\varrho$
but does not depend explicitly on $\mathbf{q}$
and the other expressed only with $\mathbf{q}$,
\begin{equation}
  E^\mathrm{KS}_\mathrm{irr}[\varrho;\mathbf{q}]
  = \sum_{k\ell} t_{k\ell}\,\varrho_{k\ell}
  + \mathit{\Delta}E^\mathrm{KS}_\mathrm{irr}[\mathbf{q}]\,,
\label{eq:Eirr-sep}\end{equation}
Eq.~\eqref{eq:KS-hamil} becomes
\begin{equation}
  h^\mathrm{KS}_{k\ell}
  = t_{k\ell} + \Gamma^\mathrm{KS}_{k\ell}\,;\quad
  \Gamma^\mathrm{KS}_{k\ell}
  = \sum_A \bigg(\frac{\partial\mathit{\Delta}E^\mathrm{KS}_\mathrm{irr}
    [\mathbf{q}]}{\partial q^A}
  + \frac{\partial E^\mathrm{KS}_\mathrm{reg}[\mathbf{q}]}{\partial q^A}\bigg)\,
  \frac{\partial Q^A}{\partial\varrho_{k\ell}}\,.
\label{eq:KS-hamil2}\end{equation}
This is the case of the standard KS equation (see Sec.~\ref{subsec:electrons}).

\subsection{Kohn-Sham--Bogolyubov-de Gennes equation\label{subsec:KSBdG-eq}}

Irregularity of $E^\mathrm{HK}[\mathbf{Q}]$
may arise via correlations linked to a pair condensate,
as in superconductivity.
It is then reasonable to include the pairing tensor $\kappa$ and $\kappa^\ast$
in the irregular part of the EDF,
and to extend Eq.~\eqref{eq:EHK-sep0} as
\begin{equation}
  E[\Psi] = E^\mathrm{KSBdG}_\mathrm{irr}\big[\varrho,\kappa,\kappa^\ast
    ;\mathbf{Q}[\varrho,\kappa,\kappa^\ast]\big]
  + E_\mathrm{reg}[\Psi]\,.
  \label{eq:EHK-sep0b}
\end{equation}
Here $Q^A$'s are postulated to depend on $(\varrho,\kappa,\kappa^\ast)$.
Arguments analogous to Eqs.~\eqref{eq:EHK-sep1} and \eqref{eq:EHK-sep2} yield
\begin{equation}\begin{split}
  & E^\mathrm{HK}[\mathbf{Q}]
  = \min_{(\varrho,\kappa,\kappa^\ast)\to\mathbf{Q}}
    E^\mathrm{KSBdG}_\mathrm{irr}[\varrho,\kappa,\kappa^\ast;\mathbf{Q}]
  + E^\mathrm{KSBdG}_\mathrm{reg}[\mathbf{Q}]\,;\\
  &\quad E^\mathrm{KSBdG}_\mathrm{reg}\big[\mathbf{Q}]
  := E_\mathrm{reg}\big[\Psi^\mathrm{HK}_\mathbf{Q}\big]
  + \Big(E^\mathrm{KSBdG}_\mathrm{irr}\big[\varrho[\Psi^\mathrm{HK}_\mathbf{Q}],
    \kappa[\Psi^\mathrm{HK}_\mathbf{Q}],
    \kappa^\ast[\Psi^\mathrm{HK}_\mathbf{Q}];\mathbf{Q}\big] \\
  &\qquad\qquad\qquad\qquad\qquad\qquad\qquad\qquad
  - \min_{(\varrho,\kappa,\kappa^\ast)\to\mathbf{Q}}
    E^\mathrm{KSBdG}_\mathrm{irr}[\varrho,\kappa,\kappa^\ast;\mathbf{Q}]
  \Big)\,.
\end{split}\label{eq:EHK-sep2b}
\end{equation}
As the KS functional in Eq.~\eqref{eq:KS-EDF},
the Kohn-Sham--Bogolyubov-de Gennes (KSBdG) functional is obtained
by regarding $q^A=Q^A$ as a constraint,
\begin{equation}
  E^\mathrm{KSBdG}[\varrho,\kappa,\kappa^\ast;\mathbf{q}] =
  E^\mathrm{KSBdG}_\mathrm{irr}[\varrho,\kappa,\kappa^\ast;\mathbf{q}]
  + E^\mathrm{KSBdG}_\mathrm{reg}[\mathbf{q}]
  - \sum_A \Lambda^\mathrm{KSBdG}_A\,\Big(q^A - Q^A[\varrho,\kappa,\kappa^\ast]\Big)\,.
  \label{eq:KBdGS-EDF}
\end{equation}
In addition, the particle-number condition \eqref{eq:N-cond} is considered,
\begin{equation}
  \tilde{E}^\mathrm{KSBdG}
  := E^\mathrm{KSBdG} - \mu \big[\mathrm{tr}(\varrho) - N\big]\,.
  \label{eq:Emod_KSBdG}
\end{equation}
By defining
\begin{equation}
  \mathsf{H}^\mathrm{KSBdG}
  := \begin{pmatrix} h^\mathrm{KSBdG}-\mu & \Delta^\mathrm{KSBdG} \\
    -\Delta^{\ast\mathrm{KSBdG}} & -(h^\mathrm{KSBdG})^\ast+\mu \end{pmatrix}\,,
\end{equation}
where
\begin{equation}\begin{split}
  h^\mathrm{KSBdG}_{k\ell}
  &:= \frac{\partial E^\mathrm{KSBdG}}{\partial\varrho_{\ell k}}
  = \frac{\partial E^\mathrm{KSBdG}_\mathrm{irr}
    [\varrho,\kappa,\kappa^\ast;\mathbf{q}]}{\partial\varrho_{\ell k}}
  + \Gamma^\mathrm{KSBdG}_{k\ell}\,;\\
  & \Gamma^\mathrm{KSBdG}_{k\ell}
  := \sum_A \Lambda^\mathrm{KSBdG}_A\,\frac{\partial Q^A}{\partial\varrho_{k\ell}}
  = \sum_A \bigg(\frac{\partial E^\mathrm{KSBdG}_\mathrm{reg}[\mathbf{q}]}
  {\partial q^A}
  + \frac{\partial E^\mathrm{KSBdG}_\mathrm{irr}
  [\varrho,\kappa,\kappa^\ast;\mathbf{q}]}{\partial q^A}\bigg)\,
  \frac{\partial Q^A}{\partial\varrho_{k\ell}}\,,\\
  \Delta^\mathrm{KSBdG}_{k\ell}
  &:= -\frac{\partial E^\mathrm{KSBdG}}{\partial\kappa^\ast_{\ell k}} \\
  &= - \frac{\partial E^\mathrm{KSBdG}_\mathrm{irr}
    [\varrho,\kappa,\kappa^\ast;\mathbf{q}]}{\partial\kappa^\ast_{\ell k}}
  - \sum_A \bigg(\frac{\partial E^\mathrm{KSBdG}_\mathrm{reg}[\mathbf{q}]}
  {\partial q^A}
  + \frac{\partial E^\mathrm{KSBdG}_\mathrm{irr}
  [\varrho,\kappa,\kappa^\ast;\mathbf{q}]}{\partial q^A}\bigg)\,
  \frac{\partial Q^A}{\partial\kappa^\ast_{k\ell}}\,,\\
  \Delta^{\ast\mathrm{KSBdG}}_{k\ell}
  &:= -\frac{\partial E^\mathrm{KSBdG}}{\partial\kappa_{\ell k}}
  = \big(\Delta^\mathrm{KSBdG}_{k\ell}\big)^\ast\,,
\end{split}\label{eq:h&D_KSBdG}
\end{equation}
the KSBdG equation is derived,
\begin{equation}
  \sum_\ell \mathsf{H}^\mathrm{KSBdG}_{k\ell}\,
  \mathsf{W}^\mathrm{KSBdG}_{\ell i}
  = \varepsilon^\mathrm{KSBdG}_i\,\mathsf{W}^\mathrm{KSBdG}_{ki}\,,
  \label{eq:KSBdG-eq}
\end{equation}
in analogy to the HFB equation \eqref{eq:HFB-eq}.
Above, the relation obtained by the variation with respect to $q^A$,
\begin{equation}
  \Lambda^\mathrm{KSBdG}_A
  = \frac{\partial E^\mathrm{KSBdG}_\mathrm{irr}
    [\varrho,\kappa,\kappa^\ast;\mathbf{q}]}{\partial q^A}
  + \frac{\partial E^\mathrm{KSBdG}_\mathrm{reg}[\mathbf{q}]}{\partial q^A}\,.
\end{equation}
has been inserted.
The identical mathematical structure guarantees
that the arguments on the solutions of the HFB equation in Appendix~\ref{app:HFB}
also apply to those of the KSBdG equation \eqref{eq:KSBdG-eq}.

It is noted that $\kappa$ and $\kappa^\ast$ do not need to be physical
in the KSBdG scheme.
The KSBdG solution can be exact for systems
where the particle number is fixed,
provided that $E^\mathrm{KSBdG}$ is appropriately determined.

\section{Discussions on principal variables
  and energy functional\label{sec:EDF}}

The reformulation of the KS theory elucidates several aspects of the theory.
Some of them are new, not recognized sufficiently,
while others have been known but can now be argued in an organized manner.
We hereafter focus on the KS equation for the sake of simplicity.
Extension to the KSBdG equation will be straightforward.

\subsection{Principal variables and density matrix\label{subsec:prince-sp}}

Whereas the HK theorem in Sec.~\ref{sec:HK} does not demand it,
all $Q^A$'s have been assumed to depend solely on $\varrho$
in Sec.~\ref{subsec:KS-eq},
not on $\mathcal{C}^{(\nu)}$ ($\nu\geq 2$).
Comments on this point are given.

The KS equation closes within the s.p. degrees of freedom,
not leading to hierarchical equations.
This remarkable property is actualized
because everything in $E^\mathrm{KS}$ does not depend
on the correlation functions $\mathcal{C}^{(\nu)}$ ($\nu\geq 2$).
The principal variables $\mathbf{Q}$ should not depend on $\mathcal{C}^{(\nu)}$
for the KS equation to keep mathematical rigor.

One might consider that the KS equation can be derived
even when $Q^A$ depends on $\mathcal{C}^{(\nu)}$
via a constrained search similar to Eq.~\eqref{eq:EHK}~\cite{Donnely-Parr_1978}.
For instance,
suppose that $Q^A$ depends on $\mathcal{C}^{(2)}$ as well as on $\varrho$.
Then $E^\mathrm{KS}$ of Eq.~\eqref{eq:KS-EDF}
also depends on $\mathcal{C}^{(2)}$.
Not to lead to hierarchical equations,
a new EDF should be constructed via the constrained search
$\min_{(\varrho,\mathcal{C}^{(2)})\to\varrho} E^\mathrm{KS}$.
However, this introduces another step in the minimization process,
which likely damages the differentiability of $Q^A$ for $\varrho$,
as illustrated in Appendix~\ref{app:2-step}.
Because the KS theory relies on this differentiability,
it is desired that the principal variables
do not depend on $\mathcal{C}^{(\nu)}$.
Conversely, if $\mathcal{C}^{(\nu)}$ is initially contained in $Q^A$
but is finally eliminated in $E^\mathrm{KS}$,
one should pay special attention to the differentiability.

\subsection{Equational equivalence\label{subsec:equivalence}}

It is worth recalling
that the KS equation \eqref{eq:KS-eq} is derived
primarily from the variation of $E^\mathrm{KS}[\varrho;\mathbf{q}]$
with respect to $\varrho$.
Apart from the physical importance,
the variation with respect to $q^A$ is subsidiary
in deriving the equational form,
only determining the KS potential $\Gamma^\mathrm{KS}$
through the Lagrange multiplier $\Lambda^\mathrm{KS}_A$
as in Eqs.~\eqref{eq:KSpot} and \eqref{eq:KS-hamil}.
Indeed, variation of $E^\mathrm{KS}$ for $\varrho$
after inserting $q^A=Q^A[\varrho]$
leads to the same s.p. Hamiltonian as in Eq.~\eqref{eq:KS-hamil},
\begin{equation}
  h^\mathrm{KS}_{k\ell} = \frac{\partial}{\partial\varrho_{k\ell}}
  E^\mathrm{KS}\big[\varrho;\mathbf{Q}[\varrho]\big]
  = \frac{\partial}{\partial\varrho_{k\ell}}
  \Big(E^\mathrm{KS}_\mathrm{irr}\big[\varrho;\mathbf{Q}[\varrho]\big]
  + E^\mathrm{KS}_\mathrm{reg}\big[\mathbf{Q}[\varrho]\big]\Big)\,.
\end{equation}
The g.s. energy is given at $\varrho=\varrho^\mathrm{KS}$,
\begin{equation}
  E_0 = E^\mathrm{KS}\big[\varrho^\mathrm{KS};
    \mathbf{Q}[\varrho^\mathrm{KS}]\big]
  = E^\mathrm{KS}_\mathrm{irr}\big[\varrho^\mathrm{KS};
    \mathbf{Q}[\varrho^\mathrm{KS}]\big]
  + E^\mathrm{KS}_\mathrm{reg}\big[\mathbf{Q}[\varrho^\mathrm{KS}]\big]\,.
\end{equation}
This feature implies that equations equivalent to \eqref{eq:KS-eq} are obtained
by removing or adding some variables to $\mathbf{Q}$,
as long as the total $E^\mathrm{KS}$ is an identical functional of $\varrho$.
Consider the case that some of the principal variables are taken out.
The new set is denoted by $\mathbf{Q}'=\{Q^{B'}\}$
and the removed set by $\bar{\mathbf{Q}}=\{Q^{\bar{B}}\}$;
$\mathbf{Q}=\mathbf{Q}'\oplus\bar{\mathbf{Q}}$.
$E^\mathrm{KS}_\mathrm{reg}$ is separated into two parts,
$E^\mathrm{KS}_\mathrm{reg}=\bar{E}^\mathrm{KS}_\mathrm{reg}
+E^{\prime\,\mathrm{KS}}_\mathrm{reg}$,
where $\bar{E}^\mathrm{KS}_\mathrm{reg}$ includes $Q^{\bar{B}}$
and $E^{\prime\,\mathrm{KS}}_\mathrm{reg}$ does not.
By newly defining $E^{\prime\,\mathrm{KS}}_\mathrm{irr}
:=E^\mathrm{KS}_\mathrm{irr}+\bar{E}^\mathrm{KS}_\mathrm{reg}$,
the following equality,
\begin{equation}
  E^{\prime\,\mathrm{KS}}_\mathrm{irr}\big[\varrho;\mathbf{Q}'[\varrho]\big]
  + E^{\prime\,\mathrm{KS}}_\mathrm{reg}\big[\mathbf{Q}'[\varrho]\big]
  = E^\mathrm{KS}_\mathrm{irr}\big[\varrho;\mathbf{Q}[\varrho]\big]
  + E^\mathrm{KS}_\mathrm{reg}\big[\mathbf{Q}[\varrho]\big]\,,
\end{equation}
is fulfilled up to the functional form of $\varrho$.
This relation guarantees the equivalence at the level of the equations
as well as the resultant energy.
The opposite is true:
principal variables may be added without influencing the s.p. equations.
We shall call this feature equational equivalence.

Despite the equational equivalence,
the choice of $\mathbf{Q}$ is physically significant.
As already discussed,
while the energy $E$ and the principal variables $\mathbf{Q}$
obtained from the KS equation are physical,
other quantities do not have strict physical meaning. 
The equational equivalence indicates
that the equational form does not tell what variables are physical.
It entirely depends on how the EDF has been constructed.
This point matters when the EDF is phenomenologically fixed
as in Sec.~\ref{subsec:nucleons}.

Several limiting cases are instructive.
First, if all $Q^A$'s are taken out by setting $\bar{\mathbf{Q}}=\mathbf{Q}$,
$\mathbf{Q}'$ becomes an empty set $\emptyset$,
leading to $E^{\prime\,\mathrm{KS}}_\mathrm{reg}=0$
and $E^{\mathrm{KS}}[\varrho;\emptyset]
=E^{\prime\,\mathrm{KS}}_\mathrm{irr}[\varrho;\emptyset]$.
In this case, no quantity other than the total energy is considered physical.
Second, it is possible to set $E^{\prime\,\mathrm{KS}}_\mathrm{irr}=0$,
opposite to the first limiting case.
There are practical cases
in which $E^\mathrm{KS}_\mathrm{irr}$ can be represented
in terms of certain currents.
For instance, in the standard KS theory for electronic systems,
the kinetic density of the non-interacting system is expressed
as $\xi(\mathbf{r})/2m$,
where
\begin{equation}
  \xi(\mathbf{r}) := \sum_{i=1}^N [\nabla\varphi_i^\ast(\mathbf{r})]\cdot
  [\nabla\varphi_i(\mathbf{r})]\,.
\end{equation}
If all those currents are added to $\mathbf{Q}$,
we have $E^{\prime\,\mathrm{KS}}_\mathrm{irr}=0$.
Still, we should consider $E^{\prime\,\mathrm{KS}}_\mathrm{irr}$
to depend on $\varrho$,
since it is essential to derive the KS equation.
An extreme case along this line
is that of $\mathbf{Q}=\{\varrho_{k\ell}\}$,
the whole 1-body density matrix~\cite{Donnely-Parr_1978,Zumbach-Maschke_1985,
  Requist-Pankratov_2008,Pernal-Giesbertz_2016}.
Then all components of $\varrho$ would be considered physical.
However, an idempotent $\varrho$,
which necessarily results from the KS equation,
cannot be exact in general,
even though there remains room to interpret the s.p. states
as those of dressed particles.
These limiting cases help to understand issues of representability,
as discussed in the subsequent subsection.

\subsection{Representability in the KS theory\label{subsec:representability}}

Significance of representability has been perceived for the HK theorem.
The HK theorem in Sec.~\ref{sec:HK} was initially asserted
by showing the uniqueness of the mapping
from the electron density $n(\mathbf{r})$
to the external electron-ion potential
$U_{ei}(\mathbf{r})$~\cite{Hohenberg-Kohn},
which are connected via the Legendre transformation.
This mapping should exist for any $n(\mathbf{r})$
with $n(\mathbf{r})\geq 0$ and $\int d^3r\,n(\mathbf{r}) = N$
(the $v$-representability).
While this $v$-representability issue can be circumvented
by the constrained search of Eq.~\eqref{eq:EHK},
it is transferred to the differentiability of the EDF
as mentioned in Sec.~\ref{sec:HK},
and this problem could arise even at the KS level.
Once $\mathbf{\Lambda}^\mathrm{KS}=\{\Lambda^\mathrm{KS}_A\}$ are fixed,
the KS equation \eqref{eq:KS-eq} with the KS Hamiltonian \eqref{eq:KS-hamil}
determines $\varrho^\mathrm{KS}$
and thereby the principal variables $\mathbf{Q}[\varrho^\mathrm{KS}]$,
defining the mapping from $\mathbf{\Lambda}^\mathrm{KS}$ to $\mathbf{Q}$.
The question is in the inverse mapping
from $\mathbf{Q}$ to $\mathbf{\Lambda}^\mathrm{KS}$,
\textit{i.e.}, whether $\mathbf{\Lambda}^\mathrm{KS}$ exists
for any $\mathbf{Q}$ in its physical domain.
The existence of this inverse mapping
may be regarded as $v$-representability at the KS level,
which has been called `non-interacting $v$-representability' in some literature.
Equation~\eqref{eq:KSpot} tells us that the $v$-representability at the KS level
arrives at the differentiability
of $E^\mathrm{KS}_\mathrm{irr}[\varrho;\mathbf{Q}]$
and $E^\mathrm{KS}_\mathrm{reg}[\mathbf{Q}]$ with respect to $Q^A$.

As already mentioned,
any $\mathbf{Q}$ in the physical domain should be represented
by the 1-body density matrix $\varrho$
that satisfies Eqs.~\eqref{eq:N-cond} and \eqref{eq:Pauli}.
This condition is called $N$-representability
and has been proven for the original case
$\mathbf{Q}=n(\mathbf{r})$~\cite{Harriman}.
For the KS theory to be applicable,
any $\mathbf{Q}$ in its physical domain
should be obtained by an idempotent $\varrho$.
This condition is stronger than the $N$-representability for the HK theorem
and the `pure-state representability' in Ref.~\cite{Valone_1980}.
We shall call it $N$-representability at the KS level.
For $\mathbf{Q}=n(\mathbf{r})$,
the $N$-representability is also satisfied at the KS level~\cite{Harriman}.

Two representabilities have been distinguished in the fermionic DFT,
the $v$- and the $N$-representabilities.
Both are further distinguished between those at the HK and KS levels.
The $v$-representability is linked to the differentiability of the EDF,
and it is usually weaker at the KS level than at the HK level.
On the other hand, the $N$-representability at the KS level,
which requires the idempotency of the 1-body density matrix,
is stronger than that at the HK level.

\subsection{Criteria for extendability\label{subsec:criterion}}

The reformulation of the KS theory with the least postulates
addresses the criteria for its extendability.
The following conditions should be satisfied
for the KS equation to be rigorous and practical,
and they give the criteria.
\begin{itemize}
\item[(a)]\label{it:Q} Principal variables $\mathbf{Q}$
  should be appropriately chosen.
  They should help to minimize energy
  and contain all quantities under physical interest other than energy.
  At the same time, all $Q^A$'s should depend only on $\varrho$
  and not on the correlation functions $\mathcal{C}^{(\nu)}$ ($\nu\geq 2$),
  or well approximated by functionals only of $\varrho$.
  Moreover, $\mathbf{Q}$ should be $N$-representable at the KS level
  in the domain under search,
  \textit{i.e.}, representable by an idempotent $\varrho$
  with $\mathrm{tr}(\varrho)=N$.
\item[(b)]\label{it:sp} Irregularity in $E^\mathrm{HK}[\mathbf{Q}]$
  can be removed via $E^\mathrm{KS}_\mathrm{irr}[\varrho;\mathbf{Q}]$,
  which is a differentiable functional of $\varrho$.
  Furthermore, $E^\mathrm{KS}_\mathrm{irr}$ should properly be chosen
  so that $E^\mathrm{KS}_\mathrm{reg}[\mathbf{Q}]$ should be
  regular for $\mathbf{Q}$.
  Because $E^\mathrm{KS}_\mathrm{reg}$ is defined by \eqref{eq:EHK-sep2},
  this condition usually requires
  that both $E^\mathrm{HK}_\mathrm{reg}[\mathbf{Q}]$
  and $\Big(E^\mathrm{KS}_\mathrm{irr}\big[\varrho[\Psi^\mathrm{HK}_\mathbf{Q}];
    \mathbf{Q}\big]
  - \min_{\varrho\to\mathbf{Q}} E^\mathrm{KS}_\mathrm{irr}[\varrho;\mathbf{Q}]
  \Big)$ are regular for $\mathbf{Q}$.
\item[(c)]\label{it:reg} The entire $E^\mathrm{KS}[\varrho;\mathbf{Q}]$
  should be fixed before starting computation,
  which requires that $E^\mathrm{KS}_\mathrm{reg}[\mathbf{Q}]$ is constructed
  appropriately.
\end{itemize}

Condition (a) is the first step for applying the KS approach.
One might think that the greater number of principal variables
provide the more accurate description of the system.
However, the $N$-representability becomes the more difficult to be satisfied.
Therefore the set $\mathbf{Q}$ should be neither too big nor too small.
Significance of (b) is exemplified
by the normal-to-superconductor transition.
The pair correlation carried by $\kappa$ and $\kappa^\ast$
plays an essential role in this phenomenon,
connected to the order parameter.
It is difficult to remove the irregularity at the transition
only by $\varrho$,
indicating that the KS theory itself cannot be applied,
though possible via the extension to the KSBdG equation\footnote{
  For the KSBdG scheme,
  $\varrho$ in Conditions (a--c) should be read
  as $(\varrho,\kappa,\kappa^\ast)$.
  }.
Another example could be the confinement of constituent quarks,
in which the 3-body correlation forming the color singlet is essential.
It does not seem possible to ascribe irregularity in $E^\mathrm{HK}$
to $E^\mathrm{KS}_\mathrm{irr}$ that is a functional only of $\varrho$,
without using $\mathcal{C}^{(3)}$.

If the underlying Hamiltonian $H$ is known and is partitioned
as in Eq.~\eqref{eq:Hamil-sep0},
it may be possible to consider $H_\mathrm{irr}$ and $H_\mathrm{reg}$
unperturbed Hamiltonian and perturbation, respectively.
Then, when setting on $H_\mathrm{reg}$ adiabatically,
the wave-function $|\Psi\rangle$ changes gradually,
as the adiabatic theorem~\cite{Messiah_2} tells us.
At a glance, it looks reasonable to expect
that $E^\mathrm{KS}_\mathrm{reg}\big[\mathbf{Q}]$ is regular in this case.
However, this is not always true
because it breaks down for the $\mathbf{Q}=\{\varrho_{k\ell}\}$ case
discussed in Sec.~\ref{subsec:equivalence},
in which the $N$-representability at the KS level is lost.

Unless all conditions (a--c) are fulfilled,
approaches using the KS equation lose firmness or practicality.
Conversely, as long as an EDF meets these conditions,
the many-body system can be solved via s.p. equations,
following the procedure in Sec.~\ref{sec:KS}.
Extensions may be adding principal variables,
modifying the EDF,
or applying to other many-fermion systems than those composed of electrons.

\section{Practice of Kohn-Sham theory\label{sec:practice}}

We next show how the above arguments on the KS theory
are related to practical cases.

\subsection{Many-electron systems:
  From atoms and molecules to solids\label{subsec:electrons}}

The KS theory has been successful for electronic systems,
from atoms and molecules to solids.
Under $\mathbf{Q}=n(\mathbf{r})$,
it has been established as a standard \textit{ab initio} method
for many-electron systems.
The Hamiltonian of the system is comprised of the kinetic energy $K$,
the external potential provided by ions $U_{ei}$,
and the electron-electron interaction $V_{ee}$.
In the context of Eq.~\eqref{eq:Hamil-sep0},
these terms are partitioned as\footnote{
Although the contribution of $U_{ei}$ is regular for $n(\mathbf{r})$,
we put it into $H_\mathrm{irr}$ so that $H_\mathrm{reg}$ could be universal.
Shifting it to $H_\mathrm{reg}$ does not influence the resultant equations.},
\begin{equation}
  H_\mathrm{irr} = K + U_{ei}\,,\quad H_\mathrm{reg} = V_{ee}\,.
  \label{eq:H_electrons}
\end{equation}
Correspondingly, the EDF follows Eq.~\eqref{eq:Hamil-sep0}.
As in Eq.~\eqref{eq:Eirr-sep},
$E^\mathrm{KS}_\mathrm{irr}$ is separated
into individual terms of Eq.~\eqref{eq:H_electrons} as
\begin{equation}\begin{split}
    E^\mathrm{KS}_\mathrm{irr}[\varrho;n(\mathbf{r})]
    &= E_K[\varrho] + E_{ei}[n(\mathbf{r})]\,;\\
    & E_K[\varrho] = -\frac{1}{2m}\sum_{k\ell}\langle\ell|\nabla^2|k\rangle\,
    \varrho_{k\ell}\,,
    \quad E_{ei}[n(\mathbf{r})]
    = \int d^3r'\,U_{ei}(\mathbf{r'})\,n(\mathbf{r'})\,,
\end{split}\label{eq:E_irr_electrons}
\end{equation}
where $E_K[\varrho]$ corresponds to the first term of Eq.~\eqref{eq:Eirr-sep}
and $E_{ei}[n(\mathbf{r})]=\mathit{\Delta}E^\mathrm{KS}_\mathrm{irr}$.
Then $E^\mathrm{KS}_\mathrm{reg}$ in Eq.~\eqref{eq:EHK-sep2} becomes
\begin{equation}
  E^\mathrm{KS}_\mathrm{reg}[n(\mathbf{r})]
  = E^\mathrm{HK}_\mathrm{reg}[n(\mathbf{r})]
  + \Big(E_K\big[\varrho[\Psi^\mathrm{HK}_{n(\mathbf{r})}]\big]
  - \min_{\varrho\to n(\mathbf{r})} E_K[\varrho]\Big)\,,
\end{equation}
having no contribution from $E_{ei}$.
Since $U_{ei}$ in Eq.~\eqref{eq:H_electrons}
carries characteristics of individual systems,
$E^\mathrm{KS}_\mathrm{reg}$ is expected to be universal.
As long as $H_\mathrm{reg}$ or its exchange-correlation part shown below
is perturbative,
the regularity of $E^\mathrm{KS}_\mathrm{reg}$ is likely satisfied.
There have been attempts to extend the theory
to include other variables in $\mathbf{Q}$
as mentioned in Sec.~\ref{sec:intro},
although the representability issues have not been explored sufficiently.

$E^\mathrm{KS}_\mathrm{reg}$ is taken
as a sum of the direct term of $V_{ee}$ and the rest:
\begin{equation}\begin{split}
    E^\mathrm{KS}_\mathrm{reg}[n(\mathbf{r})]
    &= E_\mathrm{dir}[n(\mathbf{r})]
    + E^\mathrm{KS}_\mathrm{xc}[n(\mathbf{r})]\,;\\
    & E_\mathrm{dir}[n(\mathbf{r})]
    = \frac{1}{2}\int d^3r'\,d^3r''\,
    \frac{n(\mathbf{r}')\,n(\mathbf{r}'')}{|\mathbf{r}'-\mathbf{r}''|}\,,
    \quad E^\mathrm{KS}_\mathrm{xc}[n(\mathbf{r})]
    = \int d^3r'\,\mathcal{H}_\mathrm{xc}[n(\mathbf{r}')]\,.
\end{split}\label{eq:E_reg_electrons}
\end{equation}
All of the complicated many-body effects are attributed
to the exchange-correlation term $E^\mathrm{KS}_\mathrm{xc}$.
It should be noted that the general theory does not demand
$E^\mathrm{KS}_\mathrm{reg}$ to be represented by an integral
of a local function.
Indeed, $E_\mathrm{dir}$ is an integral containing the densities
at distant positions $\mathbf{r}'$ and $\mathbf{r}''$,
multiplied by the interaction
having the non-local form $1/|\mathbf{r}'-\mathbf{r}''|$.
In contrast, $E^\mathrm{KS}_\mathrm{xc}$ is customarily assumed
to be the integral of a local function $\mathcal{H}_\mathrm{xc}$,
a function depending on the density at a single position $\mathbf{r}'$.
The kernel $\mathcal{H}_\mathrm{xc}$ has been determined
by the local-density approximation (LDA)~\cite{
  Hohenberg-Kohn,Sahni-Bohnen-Harbola}
or the generalized gradient approximation (GGA)~\cite{
  Perdew-Burke-Ernzerhof_1996}.
A method for obtaining $\mathcal{H}_\mathrm{xc}$ called `meta-GGA'
has been developed~\cite{Tao-Perdew-Staroverov-Scuseria_2003}.
A machine-learning-based method has been proposed
recently~\cite{Nagai-Akashi-Sugino_2022}.
While the locality assumption on $E^\mathrm{KS}_\mathrm{xc}$
has helped the KS theory to be practical in electronic systems,
it is an approximation whose appropriateness should be scrutinized.

It deserves discussing the spin DFT in the absence of external magnetic field,
which may exemplify arguments in Sec.~\ref{sec:EDF}.
In the spin DFT,
it is usually considered that the spin density
$\sigma(\mathbf{r}):=n_\uparrow(\mathbf{r})-n_\downarrow(\mathbf{r})$
is the principal variable additional to the density
$n(\mathbf{r})=n_\uparrow(\mathbf{r})+n_\downarrow(\mathbf{r})$.
The spin density $\sigma(\mathbf{r})$ appears in the exchange-correlation term
in the KS EDF as
\begin{equation}
    E^\mathrm{KS}[\varrho;n(\mathbf{r}),\sigma(\mathbf{r})]
    = E_K[\varrho] + E_{ei}[n(\mathbf{r})] + E_\mathrm{dir}[n(\mathbf{r})]
    + E^\mathrm{KS}_\mathrm{xc}[n(\mathbf{r}),\sigma(\mathbf{r})]\,.
\label{eq:KS-EDF_spin}
\end{equation}
It should be kept in mind that the $N$-representability has not been proven
for the case $\mathbf{Q}=\{n(\mathbf{r}),\sigma(\mathbf{r})\}$.

The system can spontaneously be magnetized.
Let us go back to the EDF at the HK level.
As far as $\sigma(\mathbf{r})/n(\mathbf{r})$ is small,
we can express the HK EDF as
\begin{equation}
  E^\mathrm{HK}[n(\mathbf{r}),\sigma(\mathbf{r})]
  = C_0^\mathrm{HK}[n(\mathbf{r})]
  + C_2^\mathrm{HK}[n(\mathbf{r})]\cdot[\sigma(\mathbf{r})]^2
  + C_4^\mathrm{HK}[n(\mathbf{r})]\cdot[\sigma(\mathbf{r})]^4\,.
\label{eq:EHK_spin}\end{equation}
We assume $C_4^\mathrm{HK}[n]>0$ everywhere,
and that $C_2^\mathrm{HK}[n]$ changes its sign
at a certain value of $n$, denoted by $n_\mathrm{cr}$:
\begin{equation}
  \left\{\begin{array}{ll} C_2^\mathrm{HK}[n] >0 &\mbox{for $n<n_\mathrm{cr}$} \\
  C_2^\mathrm{HK}[n] < 0 &\mbox{for $n>n_\mathrm{cr}$} \end{array}\right.\,.
\end{equation}
Regarding Eq.~\eqref{eq:KS-EDF_spin},
the $C_2$ and $C_4$ terms mainly correspond to $E^\mathrm{KS}_\mathrm{xc}$,
although $E_K$ has some contribution due to the Pauli principle,
which should be positive for $C_2$.
Minimizing $E^\mathrm{HK}[n(\mathbf{r}),\sigma(\mathbf{r})]$
with respect to $\sigma$,
we find a non-magnetized phase and a magnetized phase,
\begin{equation}\begin{split}
 & \min_{\sigma(\mathbf{r})} E^\mathrm{HK}[n(\mathbf{r}),\sigma(\mathbf{r})] \\
 &\quad = \left\{\begin{array}{lll}
    C_0^\mathrm{HK}[n(\mathbf{r})] &\mbox{with } \sigma(\mathbf{r})=0
    &\mbox{(for $n(\mathbf{r})<n_\mathrm{cr}$)} \\
    {\displaystyle C_0^\mathrm{HK}[n(\mathbf{r})]
      - \frac{\big|C_2^\mathrm{HK}[n(\mathbf{r})]\big|^2}
      {4\,C_4^\mathrm{HK}[n(\mathbf{r})]}}
    &\mbox{with }{\displaystyle [\sigma(\mathbf{r})]^2
      =\frac{\big|C_2^\mathrm{HK}[n(\mathbf{r})]\big|}
      {2\,C_4^\mathrm{HK}[n(\mathbf{r})]}}
    &\mbox{(for $n(\mathbf{r})>n_\mathrm{cr}$)} \end{array}\right.\,.
\end{split}\label{eq:spin-trans}\end{equation}
This spontaneous magnetization across $n=n_\mathrm{cr}$
can be handled in the KS EDF of Eq.~\eqref{eq:KS-EDF_spin}.

Since the principal variables $\mathbf{Q}=\{n(\mathbf{r}),\sigma(\mathbf{r})\}$
covers those of the ordinary KS DFT, $\mathbf{Q}'=\{n(\mathbf{r})\}$,
the EDF of the ordinary KS theory should be derived from the spin DFT,
in principle.
However, an irregularity enters
owing to the transition of the phases across $n=n_\mathrm{cr}$
in Eq.~\eqref{eq:spin-trans},
which is distinguished from the irregularity due to the Pauli principle
of Eq.~\eqref{eq:Pauli}.
In practice, the HK EDF for $\mathbf{Q}'=\{n(\mathbf{r})\}$ should be
\begin{equation}
  E^\mathrm{HK}[n(\mathbf{r})]
  = \min_{\sigma(\mathbf{r})} E^\mathrm{HK}[n(\mathbf{r}),\sigma(\mathbf{r})]
  = C_0^\mathrm{HK}[n(\mathbf{r})]
  - \delta\big(n(\mathbf{r})-n_\mathrm{cr}\big)\cdot
  \frac{\big|C_2^\mathrm{HK}[n(\mathbf{r})]\big|^2}
       {4\,C_4^\mathrm{HK}[n(\mathbf{r})]}\,.
\label{eq:EHK_spin-ord}\end{equation}
The second term of the right-hand side (rhs) implies irregularity
with respect to $n(\mathbf{r})$.
It is difficult to remove the irregularity
via $E^\mathrm{KS}_\mathrm{irr}$ of Eq.~\eqref{eq:E_irr_electrons}.
Instead, we may construct a KS EDF
by taking $E^\mathrm{KS}_\mathrm{irr}$ as
\begin{equation}
    E^\mathrm{KS}_\mathrm{irr}[\varrho;n(\mathbf{r})]
    = E_K[\varrho] + E_{ei}[n(\mathbf{r})]
  + C_2^\mathrm{KS}[n(\mathbf{r})]\cdot[\sigma(\mathbf{r})]^2
  + C_4^\mathrm{KS}[n(\mathbf{r})]\cdot[\sigma(\mathbf{r})]^4\,,
\end{equation}
where $C_\nu^\mathrm{KS}$ ($\nu=0,2,4$) is the $\sigma^\nu$ term
in $E^\mathrm{KS}_\mathrm{xc}[n,\sigma]$ of Eq.~\eqref{eq:KS-EDF_spin}.
The regular part becomes
\begin{equation}
  E^\mathrm{KS}_\mathrm{reg}[n(\mathbf{r})]
  = E_\mathrm{dir}[n(\mathbf{r})] + C_0^\mathrm{KS}[n(\mathbf{r})]\,.
\end{equation}
While this EDF well takes account of the irregularity due to the magnetization,
$\sigma(\mathbf{r})$ is not regarded as a principal variable.
This reminds us of the problem;
$N$-representability has not been proven for $\sigma(\mathbf{r})$.
We may reinterpret the spin DFT,
abandoning to keep the whole $\sigma(\mathbf{r})$ as principal variables
but regarding the magnetization $\mathcal{M}:=\int d^3r\,\sigma(\mathbf{r})$
as a principal variable.
The KS EDF of Eq.~\eqref{eq:KS-EDF_spin} is identified
as $E^\mathrm{KS}[\varrho;n(\mathbf{r}),\mathcal{M}]$,
without impairing the $N$-representability\footnote{
  The argument in Ref.~\cite{Harriman} applies
  by adopting the s.p. states $\sqrt{1-c^2}\,\phi_\uparrow(\mathbf{r})
  + c\,\phi_\downarrow(\mathbf{r})$ with $2|c|^2=1-\mathcal{M}/N$.
  }.

\subsection{Many-nucleon systems: Atomic nuclei\label{subsec:nucleons}}

The success of the KS theory in many-electron systems
has made us wonder whether an \textit{ab initio} method
for nuclear structure calculations could be developed
analogously~\cite{Drut-Furnstahl-Platter_2010}.
However, problems remain to be resolved
to establish the KS theory in nuclei.

A nucleus consists of two species of particles with nearly equal masses,
protons ($p$) and neutrons ($n$).
The particle number of each particle type, $N_\tau$ $(\tau=p,n)$, is conserved.
Equation~\eqref{eq:Emod_KS} is modified accordingly,
\begin{equation}
  \tilde{E}^\mathrm{KS} := E^\mathrm{KS}
  - \sum_{\tau=p,n} \mu_\tau \big[\mathrm{tr}_\tau(\varrho) - N_\tau\big]\,,
  \label{eq:Emod_KS_2}
\end{equation}
where $\mathrm{tr}_\tau$ is the trace over the s.p. space
belonging to the particle type $\tau$.
When the local density is taken as a principal variable,
it seems reasonable to use $n_\tau(\mathbf{r})$,
the density of each particle type.

In self-bound finite systems like atomic nuclei,
treatment of the center-of-mass (c.m.) motion is a subtle problem.
It seems desirable to construct a theory
using the internal density~\cite{Engel_2007},
$\tilde{n}_\tau(\mathbf{r}):=\langle\Psi|
\sum_{i\in\tau} \delta\big(\mathbf{r}-(\mathbf{r}_i-\mathbf{R})\big)
|\Psi\rangle$,
where $\mathbf{R}$ is the c.m. coordinate.
However, $\mathbf{R}$ in the $\delta$-function makes
$\tilde{n}_\tau(\mathbf{r})$ a many-body quantity~\cite{Nakada_2020},
prohibiting the KS equation from being applied without losing exactness,
relevantly to Condition (a) in Sec.~\ref{subsec:criterion}.
See also the arguments in Sec.~\ref{subsec:prince-sp}.
A rational approximation on $\tilde{n}_\tau(\mathbf{r})$
should be introduced to apply the KS theory.
While discussions in this line have been found
in Refs.~\cite{Barnea_2007,Messud-Bender-Suraud_2009,Messud_2013},
it has not fully been resolved,
with no well-established practical method.
Although this problem is critical in exploiting authentic DFT in nuclei,
we proceed without touching on this problem further,
leaving it for future works
and simply assuming
$\tilde{n}_\tau(\mathbf{r})\approx n_\tau(\mathbf{r})
=\langle\Psi|\sum_{i\in\tau} \delta(\mathbf{r}-\mathbf{r}_i)|\Psi\rangle$,
for the time being.

While approaches analogous to the KS theory
have widely been applied to atomic nuclei,
proper means to determine the EDFs have yet to be established.
Although the density-matrix expansion was proposed
from a microscopic standpoint~\cite{Negele-Vautherin_1972,Campi-Sprung_1972},
it has not been applied extensively.
Another way is fitting parameters in an EDF or an effective interaction
to the experimental data~\cite{UNEDF1,UNEDF2}.
We here give a typical form of the nuclear EDF,
which is called Skyrme EDF:
\begin{equation}
    E_\mathrm{Skyrme} = \int d^3r\,\mathcal{H}_\mathrm{Skyrme}
    [n_\tau(\mathbf{r}),\xi_\tau(\mathbf{r}),\boldsymbol{\zeta}_\tau(\mathbf{r})]\,,
\label{eq:Skyrme}\end{equation}
where
\begin{equation}
    n_\tau(\mathbf{r}) = \sum_{i\in\tau} \varphi_i^\ast(\mathbf{r})\,
    \varphi_i(\mathbf{r})\,,\qquad
    \xi_\tau(\mathbf{r}) = \sum_{i\in\tau}
    \nabla\varphi_i^\ast(\mathbf{r})\cdot\nabla\varphi_i(\mathbf{r})\,,\quad
    \boldsymbol{\zeta}_\tau(\mathbf{r}) = i \sum_{i\in\tau}
    \varphi_i^\ast(\mathbf{r})\,\boldsymbol{\sigma}\times\nabla
    \varphi_i(\mathbf{r})\,,
\end{equation}
and the index $i$ in the summation runs over the occupied s.p. states.
The parameters contained in $\mathcal{H}_\mathrm{Skyrme}$
are usually determined by fitting to experimental data
on some physical quantities.
As pointed out in Sec.~\ref{subsec:equivalence},
the functional form does not fix principal variables.
$E^\mathrm{KS}_\mathrm{irr}$ and $E^\mathrm{KS}_\mathrm{reg}$
are not well discriminated in $E_\mathrm{Skyrme}$. 
For instance, we have a term with $n_\tau(\mathbf{r})\,\xi_\tau(\mathbf{r})$
in $\mathcal{H}_\mathrm{Skyrme}$,
which emerges by expanding a finite-range interaction in terms of
momentum~\cite{Vautherin-Brink,Negele-Vautherin_1972},
not closely connected to the shell effects.
However, if we classify this term into $E^\mathrm{KS}_\mathrm{reg}$,
it is reasonable to consider
$\xi_\tau(\mathbf{r})$ to be physical and belong to principal variables. 
Then the kinetic term becomes physical,
which is unacceptable
because the true nuclear wave-functions have high-momentum components
beyond $\mathcal{V}_\mathrm{idem}$~\cite{Mahaux-etal_1985}.
It is crucial in this respect what quantities are employed for the fitting.
Only the quantities in the fitting protocol
can be considered principal variables,
and physical meaning is questionable for other variables.
It can be doubted that even the density $n_\tau(\mathbf{r})$ is truly physical
when not involved in the fitting protocol.

As microscopic approaches,
the density-matrix expansion has been revived
in connection to the nucleonic interaction
derived from the chiral perturbation theory~\cite{Kaiser-Fritsch-Weise_2003,
  NavarroPerez-etal_2018}.
Derivation of covariant EDFs based on the Brueckner theory
has been advanced~\cite{Shen-etal_2019}.
We also mention a few attempts to compromise
microscopic and phenomenological ways
to determine the EDF~\cite{Nakada_2020,Baldo-Schuck-Vinas_2008}.

The total angular momentum $J$ should be a good quantum number in nuclei
because of the rotational symmetry.
It is known that any even-even nucleus has $J=0$ at its g.s. with no exception.
Then the density should be rotationally symmetric,
$n_\tau(\mathbf{r})=n_\tau(r)$ where $r=|\mathbf{r}|$.
However, a number of nuclei are considered
to have deformed intrinsic states~\cite{Bohr-Mottelson_2}.
The g.s. of an even-even deformed nucleus has $J=0$
as a superposition of the intrinsic states rotated with various angles.
It is difficult to represent a g.s. of a deformed nucleus
within $\mathcal{V}_\mathrm{idem}$,
viz., with a state comprised of the KS orbitals.
In other words, the $N$-representability at the KS level is questionable.
It was argued in Ref.~\cite{Giraud-Jennings-Barrett_2008}
that the KS solution would provide an intrinsic state
rather than the observed g.s.
Though practical,
this argument seems to deviate from the variational principle
of Eq.~\eqref{eq:min_E}
and the HK theorem of Eq.~\eqref{eq:min-EHK}.
In this respect, the nuclear deformation could be an example
that Condition (b) in Sec.~\ref{subsec:criterion} does not rigorously hold.
This is another problem that has not been resolved entirely.
The angular-momentum projection~\cite{Ring-Schuck} has been developed
to restore the rotational symmetry
and applied to the solutions of Eq.~\eqref{eq:HF-eq} or \eqref{eq:KS-eq}.
The projection has also been discussed
in the context of the KS theory~\cite{Giraud-Jennings-Barrett_2008}.

\section{Summary\label{sec:summary}}

The Kohn-Sham (KS) theory provides an attractive and practical platform
for \textit{ab initio} calculations of many-fermion systems,
in which a many-fermion problem is reduced
to single-particle (s.p.) equations.
Intending to expose the minimal composition
and supply a guide for extensive application,
we have generalized and reformulated the KS theory
through the 1-body density matrix $\varrho_{k\ell}$,
without specifying principal variables for the energy minimization.
Taking into account the pairing tensor,
we have also mentioned the Kohn-Sham--Bogolyubov-de Gennes equation.

Whereas the two-step optimization of energy under Levy's constrained search
assures that the Hohenberg-Kohn (HK) theorem holds
with a universal energy density functional (EDF),
it hampers the differentiability of the EDF
and motivates the KS procedure.
The KS theory is derived by separating the EDF
into regular and irregular parts.
While the irregular part is not differentiable for the principal variables,
the differentiability for the 1-body density matrix $\varrho$ is assumed.
It is noted that the density matrix $\varrho$ itself
is not and should not be a principal variable,
introduced only to remove the irregularity in the EDF.

The issues of representability are readdressed in this context.
The $v$- and $N$-representabilities are distinguished
between the HK and KS (or the non-interacting representability) levels.
The $v$-representability at the KS level is reduced
to the differentiability with respect to the principal variables
for the regular part of the EDF.
Since the KS theory relies on variation,
differentiability is crucial to justify it.
The solution of the KS equation gives idempotent $\varrho$,
which discriminates the $N$-representabilities between the HK and KS levels.
The equational equivalence has been pointed out
before and after moving some variables from principal to subsidiary,
affecting their physical significance.
We have also discussed how these issues are related to practical cases.
The minimal composition clarifies criteria for the extendability
of the KS theory,
which will guide further development
of an \textit{ab initio} density-functional theory
including other systems, \textit{e.g.}, atomic nuclei.

\begin{acknowledgments}
The author is grateful to T.~Kotani for the discussions.
\end{acknowledgments}

\appendix

\section{Hartree-Fock-Bogolyubov approximation\label{app:HFB}}

In this Appendix,
we consider the case in which $|\Psi\rangle$ does not necessarily have
a fixed particle number,
while the expectation value is maintained via Eq.~\eqref{eq:N-cond}.
By taking into account the contribution of the pairing tensor
in addition to $\varrho$,
\begin{equation}
  \kappa_{k\ell}:=\langle\Psi|a_\ell a_k|\Psi\rangle\,,\quad
  \kappa_{k\ell}^\ast=\langle\Psi|a^\dagger_k a^\dagger_\ell|\Psi\rangle\,,
  \label{eq:pair-tensor}
\end{equation}
the energy is approximated as
$E[\Psi]\approx E^\mathrm{HFB}[\varrho,\kappa,\kappa^\ast]$,
which derives the Hartree-Fock-Bogolyubov (HFB) approximation.
The independent variables are $\varrho_{k\ell}$ for all $(k,\ell)$,
$\kappa_{k\ell}$ and $\kappa^\ast_{k\ell}$ with $k<\ell$.

By modifying the energy functional analogously to Eq.~\eqref{eq:Emod_HF},
\begin{equation}
  \tilde{E}^\mathrm{HFB}
  := E^\mathrm{HFB} - \mu \big[\mathrm{tr}(\varrho) - N\big]\,,
  \label{eq:Emod_HFB}
\end{equation}
variation yields
\begin{equation}\begin{split}
  \delta\tilde{E}^\mathrm{HFB}
  &= \sum_{k\ell} \frac{\partial E^\mathrm{HFB}}{\partial\varrho_{\ell k}}\,
  \delta\varrho_{\ell k}
  + \sum_{k>\ell} \Big(
  \frac{\partial E^\mathrm{HFB}}{\partial\kappa_{\ell k}}\,
  \delta\kappa_{\ell k}
  + \frac{\partial E^\mathrm{HFB}}{\partial\kappa^\ast_{\ell k}}\,
  \delta\kappa^\ast_{\ell k} \Big)
  - \mu \sum_k \delta\varrho_{kk} \\
  &\quad - \delta\mu \big[\mathrm{tr}(\varrho) - N\big] \\
  &= \mathrm{tr}\Big[(h^\mathrm{HFB}-\mu)\,\delta\varrho
    - \frac{1}{2}\big(\Delta^{\ast\mathrm{HFB}}\,\delta\kappa
    + \Delta^\mathrm{HFB}\,\delta\kappa^\ast\big) \Big]
  - \delta\mu \big[\mathrm{tr}(\varrho) - N\big]\,,
\end{split}\label{eq:var_E_HFB}
\end{equation}
where
\begin{equation}\begin{split}
  h^\mathrm{HFB}_{k\ell}
  := \frac{\partial E^\mathrm{HFB}}{\partial\varrho_{\ell k}}
  = \big(h^\mathrm{HFB}_{\ell k}\big)^\ast\,,\quad
  & \Delta^\mathrm{HFB}_{k\ell}
  := -\frac{\partial E^\mathrm{HFB}}{\partial\kappa^\ast_{\ell k}}
  = -\Delta^\mathrm{HFB}_{\ell k}\,,\\
  & \Delta^{\ast\mathrm{HFB}}_{k\ell}
  := -\frac{\partial E^\mathrm{HFB}}{\partial\kappa_{\ell k}}
  = \big(\Delta^\mathrm{HFB}_{k\ell}\big)^\ast\,.
\end{split}\label{eq:h&D_HFB}
\end{equation}
By doubling the matrix dimension,
the HFB Hamiltonian and the generalized density matrix are defined by~\cite{
  Ring-Schuck}
\begin{equation}
  \mathsf{H}^\mathrm{HFB}
  := \begin{pmatrix} h^\mathrm{HFB}-\mu & \Delta^\mathrm{HFB} \\
    -\Delta^{\ast\mathrm{HFB}} & -(h^\mathrm{HFB})^\ast+\mu \end{pmatrix}\,,
\end{equation}
and
\begin{equation}
  \mathsf{R}
  := \begin{pmatrix} \varrho & \kappa \\
    -\kappa^\ast & 1-\varrho^\ast \end{pmatrix}
  = \begin{pmatrix} \langle\Psi|a^\dagger_\ell a_k|\Psi\rangle &
    \langle\Psi|a_\ell a_k|\Psi\rangle \\
    \langle\Psi|a^\dagger_\ell a^\dagger_k|\Psi\rangle &
    \langle\Psi|a_\ell a^\dagger_k|\Psi\rangle \end{pmatrix}\,,
  \quad\mbox{viz.,}\quad
  \mathsf{R}^T
  = \big\langle\Psi\big|\begin{pmatrix} a^\dagger_k \\ a_k \end{pmatrix}
  \begin{pmatrix} a_\ell & a^\dagger_\ell \end{pmatrix}\big|\Psi\big\rangle\,.
  \label{eq:gen-DM}
\end{equation}
Equation~\eqref{eq:var_E_HFB} is rewritten as
\begin{equation}
  \delta\tilde{E}^\mathrm{HFB}
  = \frac{1}{2}\,\mathrm{tr}'\big(\mathsf{H}^\mathrm{HFB}\,
  \delta\mathsf{R}\big)
  - \delta\mu \big[\mathrm{tr}(\varrho) - N\big]\,.
  \label{eq:var_E_HFB-2}
\end{equation}
The expression $\mathrm{tr}'$ in the first term of the rhs
stands for the trace in the space of $\mathsf{H}^\mathrm{HFB}$
and $\mathsf{R}$,
whose dimension is twice the dimension of the s.p. space.
The diagonalization of $\mathsf{H}^\mathrm{HFB}$
via an appropriate unitary transformation $\mathsf{W}^\mathrm{HFB}$,
\begin{equation}
  \sum_\ell \mathsf{H}^\mathrm{HFB}_{k\ell}\,
  \mathsf{W}^\mathrm{HFB}_{\ell i}
  = \varepsilon^\mathrm{HFB}_i\,\mathsf{W}^\mathrm{HFB}_{ki}\,,
  \label{eq:HFB-eq}
\end{equation}
is the HFB equation.
With denoting the vector composing $\mathsf{W}^\mathrm{HFB}_{ki}$
by $\mathbf{w}^\mathrm{HFB}_i$,
Eq.~\eqref{eq:HFB-eq} is expressed as
\begin{equation}
  \mathsf{H}^\mathrm{HFB}\,\mathbf{w}^\mathrm{HFB}_i
  = \varepsilon^\mathrm{HFB}_i\,\mathbf{w}^\mathrm{HFB}_i\,.
  \label{eq:HFB-eq2}
\end{equation}
A noteworthy property of $\mathsf{H}^\mathrm{HFB}$ is
\begin{equation}
  \Sigma_x\,\mathsf{H}^\mathrm{HFB}\,\Sigma_x
  = -(\mathsf{H}^\mathrm{HFB})^\ast\,,\quad
  \Sigma_x := \begin{pmatrix} 0 & 1 \\ 1 & 0 \end{pmatrix}\,.
  \label{eq:sym-H_HFB}
\end{equation}
Equation~\eqref{eq:sym-H_HFB} ensures that,
if an eigenvalue and associated eigenvector
$(\varepsilon_i,\mathbf{w}_i)$ is a solution of Eq.~\eqref{eq:HFB-eq2},
$(-\varepsilon_i,\Sigma_x\mathbf{w}_i^\ast)$
satisfies Eq.~\eqref{eq:HFB-eq2} as well,
similarly to the RPA solution~\cite{Nakada_2016}.
Therefore, it is sufficient to solve Eq.~\eqref{eq:HFB-eq2}
for $\varepsilon_i\geq 0$.
%
Application of the unitary transformation $\mathsf{W}$
to $\mathsf{R}$ yields
\begin{equation}
  \mathsf{R}^T\quad\rightarrow\quad
  \mathsf{W}^T\mathsf{R}^T\mathsf{W}^\ast
  = \big\langle\Psi\big|\mathsf{W}^T
  \begin{pmatrix} a^\dagger_k \\ a_k \end{pmatrix}
  \begin{pmatrix} a_\ell & a^\dagger_\ell \end{pmatrix} \mathsf{W}^\ast
  \big|\Psi\big\rangle\,.
  \label{eq:Bogtr_gen-DM}
\end{equation}
The unitary transformation $\mathsf{W}$ gives a Bogolyubov transformation.
Quasiparticle operators are defined by
\begin{equation}
  \begin{pmatrix} \alpha^\dagger_i \\ \alpha_i \end{pmatrix}
  = \mathsf{W}^T \begin{pmatrix} a^\dagger_k \\ a_k \end{pmatrix}\,.
  \label{eq:Bogtr}
\end{equation}
Because the transformed $\mathsf{R}$ is
\begin{equation}
  \mathsf{W}^\dagger\mathsf{R}\mathsf{W}
  = \begin{pmatrix} \langle\Psi|\alpha^\dagger_j \alpha_i|\Psi\rangle &
    \langle\Psi|\alpha_j \alpha_i|\Psi\rangle \\
    \langle\Psi|\alpha^\dagger_j \alpha^\dagger_i|\Psi\rangle &
    \langle\Psi|\alpha_j \alpha^\dagger_i|\Psi\rangle \end{pmatrix}\,,
\end{equation}
diagonal elements in the quasiparticle representation $\mathsf{R}_{ii}
=\langle\Psi|\alpha^\dagger_i \alpha_i|\Psi\rangle$
or $\langle\Psi|\alpha_i \alpha^\dagger_i|\Psi\rangle$
satisfy $0\leq\mathsf{R}_{ii}\leq 1$,
owing to the Pauli principle.

The minimum of $E^\mathrm{HFB}$ under Constraint \eqref{eq:N-cond}
needs $\delta\tilde{E}^\mathrm{HFB}\geq 0$.
The associated wave-function is denoted by $|\Phi^\mathrm{HFB}\rangle$,
which is constituted by the solutions of Eq.~\eqref{eq:HFB-eq}.
In the vicinity of its solution,
$\delta\tilde{E}^\mathrm{HFB}$ is written as
\begin{equation}
  \delta\tilde{E}^\mathrm{HFB}
  = \sum_i \varepsilon^\mathrm{HFB}_i\,\delta\mathsf{R}_{ii}\,,
  \label{eq:var_E_HFB-3}
\end{equation}
where the summation runs over $i$ having $\varepsilon^\mathrm{HFB}_i>0$.
Because $\varepsilon^\mathrm{HFB}_i>0$ and $0\leq\mathsf{R}_{ii}\leq 1$,
an argument similar to Eqs.~\eqref{eq:var_1bd_HF} and \eqref{eq:1bd_HF}
addresses a conclusion that the HFB solution has $\mathsf{R}_{ii}=0$
(for $\varepsilon^\mathrm{HFB}_i>0$),
namely $\langle\Phi^\mathrm{HFB}|\alpha^\dagger_i\alpha_i
|\Phi^\mathrm{HFB}\rangle=0$.
Thus $|\Phi^\mathrm{HFB}\rangle$ is the vacuum of quasiparticles
defined by Eq.~\eqref{eq:Bogtr} with $\mathsf{W}=\mathsf{W}^\mathrm{HFB}$.
This is characterized by $(\mathsf{R}^\mathrm{HFB})^2=\mathsf{R}^\mathrm{HFB}$,
where $\mathsf{R}^\mathrm{HFB}:=\mathsf{R}[\Phi^\mathrm{HFB}]$.

As the vacuum of the quasiparticles,
$|\Phi^\mathrm{HFB}\rangle$ is a pair condensate,
\begin{equation}
  |\Phi^\mathrm{HFB}\rangle \propto e^\mathcal{Z}|0\rangle\,,\quad
  \mathcal{Z}=\frac{1}{2}\sum_{k\ell} z_{k\ell}\,a^\dagger_k a^\dagger_\ell\,,
  \label{eq:pair-condensate}
\end{equation}
where $|0\rangle$ is the vacuum of the particles:
$a_k|0\rangle=0$ for any $k$.
Even for systems with particle-number conservation,
the HFB approximation is valid
when the pair correlation is significant
while fluctuation of the particle number in $|\Phi^\mathrm{HFB}\rangle$
does not influence seriously.
In $|\Phi^\mathrm{HFB}\rangle$,
the component having $N\,(=\mathrm{even})$ particles is
\begin{equation}
  |\Phi_N\rangle \propto \mathcal{Z}^{N/2}|0\rangle\,.
\end{equation}
The pairing tensor of Eq.~\eqref{eq:pair-tensor} could read as
\begin{equation}
  \kappa_{k\ell}\approx\langle\Phi_{N-2}|a_\ell a_k|\Phi_N\rangle\,.
  \label{eq:pair-tensor-2}
\end{equation}
The energy functional of the form $E^\mathrm{HFB}[\varrho,\kappa,\kappa^\ast]$
follows
if $\mathcal{C}^{(2)}_{k\ell k'\ell'}\approx \kappa_{k'\ell'}^\ast \kappa_{k\ell}$
and $\mathcal{C}^{(\nu)}$'s are negligible for $\nu\geq 3$.

\section{Two-step minimization and differentiability\label{app:2-step}}

This Appendix gives an elementary example of two-step minimization
for a pedagogical purpose,
illustrating how the constrained search damages differentiability.

\begin{figure}
\centerline{\includegraphics[scale=0.5]{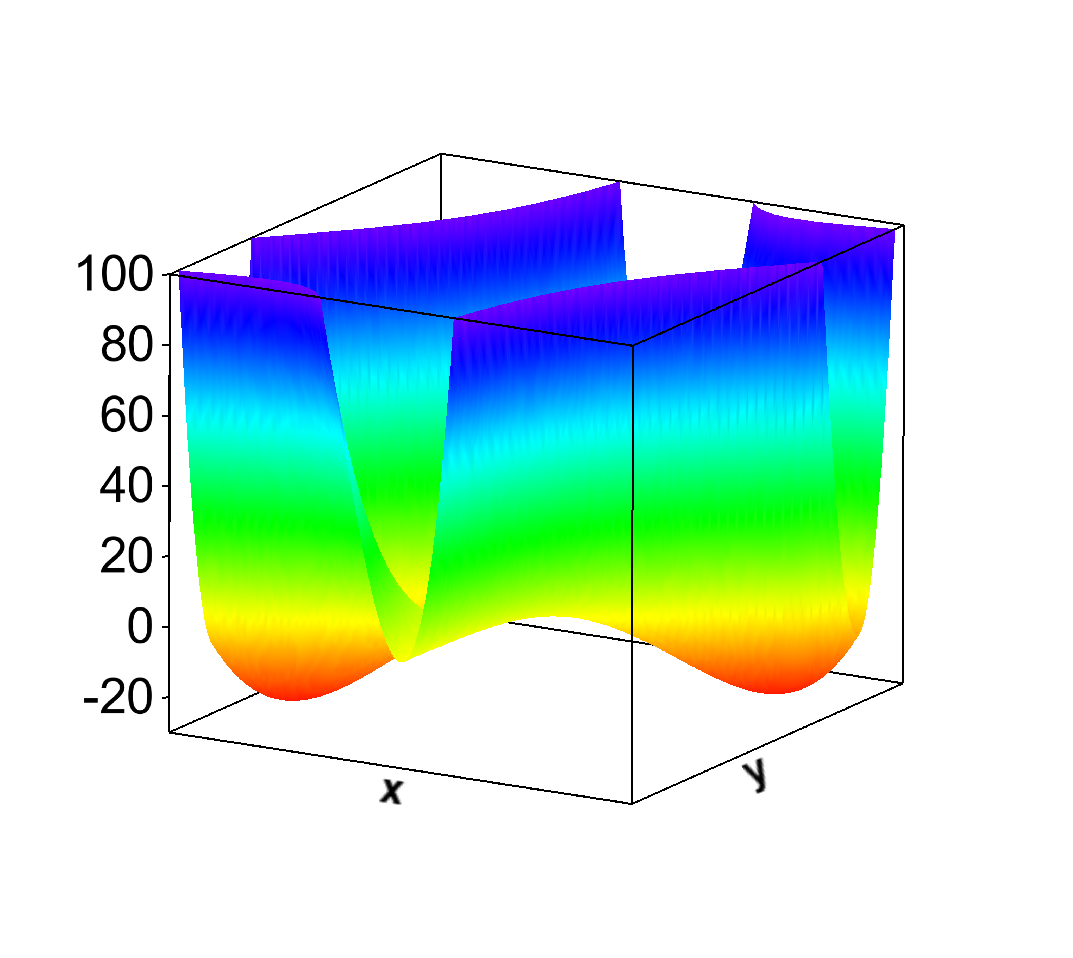}}
\caption{Landscape of $\phi(x,y)$ defined in Eq.~\eqref{eq:phi},
  in the region satisfying $-5\leq x/d\leq 5$ and $-5\leq y/d\leq 5$,
  in the unit of $d^4$.
\label{fig:phi_3D}}
\end{figure}

Consider a function $\phi(x,y)$ of the two variables $x$ and $y$,
\begin{equation}
  \phi(x,y) = 3x^4 - 8x^3y + 6x^2(y^2-d^2)\,.\quad(d>0) \label{eq:phi}
\end{equation}
This $\phi(x,y)$ is obviously analytical everywhere on the $xy$-plane.
The three-dimensional landscape of $\phi(x,y)$ is presented
in Fig.~\ref{fig:phi_3D}.
The following coupled equations give the extrema of $\phi$,
\begin{equation}\begin{split}
    \frac{\partial\phi}{\partial x}
    &= 12x(x-y+d)(x-y-d) = 0\,, \\
    \frac{\partial\phi}{\partial y}
    &= -4x^2(2x-3y) = 0\,.
\end{split}\label{eq:der_phi}\end{equation}
The solutions are
\begin{equation}
  (x,y) = (0,\mbox{arbitrary})\,,~(+3d,+2d)\,,~(-3d,-2d)\,.
\end{equation}
By substituting these solutions and comparing the values of $\phi$,
one immediately obtains
\begin{equation}
  \min_{(x,y)}\phi(x,y) = -27 d^4\quad
  \mbox{at}~(x,y)=(+3d,+2d)~\mbox{and}~(-3d,-2d)\,.
\label{eq:min_phi}\end{equation}

\begin{figure}
\centerline{\includegraphics[scale=0.6]{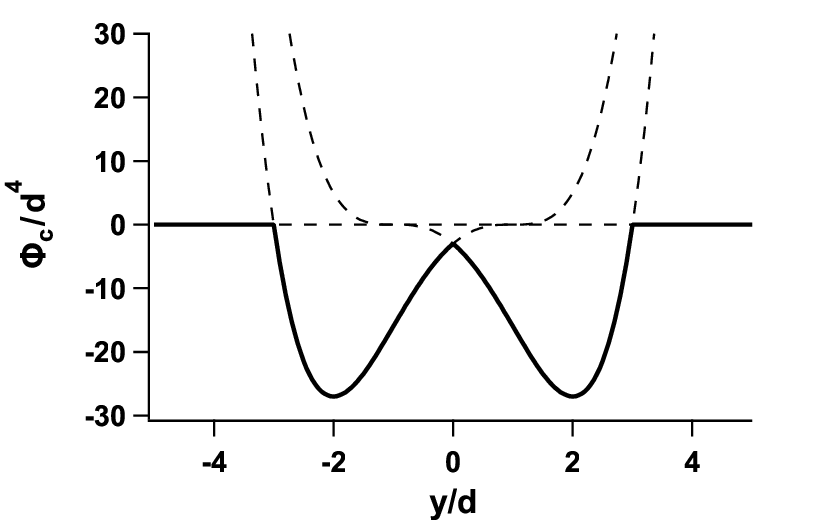}}
\caption{$\phi_c(y)$ of Eq.~\eqref{eq:phi_c} (solid line),
  and $\phi_{c0}(y),\phi_{c\pm}(y)$ of Eq.~\eqref{eq:phi_cx} (dashed lines).
\label{fig:phi_c}}
\end{figure}

The next task is to find the minimum by a two-step procedure.
First, $\phi(x,y)$ is minimized $\phi(x,y)$ with respect to $x$
for individual $y$, yielding a function $\phi_c(y)$,
\begin{equation}
  \phi_c(y) := \min_x \phi(x,y)\,.
\label{eq:def_phi_c}\end{equation}
This minimization is accomplished
via the first equation of Eq.~\eqref{eq:der_phi},
\begin{equation}
    \phi_c(y) = \min\big[\phi_{c0}(y),\phi_{c+}(y),\phi_{c-}(y)\big]
    = \left\{\begin{aligned}
    \phi_{c0}(y) &\quad(\mbox{for}~y\leq -3d~\mbox{and}~y\geq 3d) \\
    \phi_{c+}(y) &\quad(\mbox{for}~0\leq y<3d) \\
    \phi_{c-}(y) &\quad(\mbox{for}~-3d<y<0)
    \end{aligned}\right.\,,
\label{eq:phi_c}\end{equation}
where
\begin{equation}
  \phi_{c0}(y) = \phi(0,y) = 0\,,\quad
  \phi_{c\pm}(y) = \phi(y\pm d,y) = (y\pm d)^3(y\mp 3d)\,.
\label{eq:phi_cx}\end{equation}
As shown in Fig.~\ref{fig:phi_c},
$\phi_c(y)$ takes a minimum at $y=\pm 2d$,
satisfying
\begin{equation}
  \min_y \phi_c(y) = \min_{(x,y)}\phi(x,y) = -27 d^4\,.
\end{equation}
However, $\phi_c(y)$ is not differentiable at $y=0$ and $\pm 3d$,
as seen in Fig.~\ref{fig:phi_c}.
Thus, the differentiability is lost in $\phi_c(y)$ at the global level,
though it could hold within limited domains, \textit{e.g.}, $0<y<3d$.
The violation of differentiability at the global level
indicates the non-existence of universal function,
as exemplified in Eq.~\eqref{eq:phi_c}.
It is problematic to apply variational procedures
without knowing the analyticity of $\phi_c(y)$,
possibly hampering practicality.

\section{Energy density functional for 1-dimensional harmonic oscillator
  \label{app:HO-EDF}}

While the original HK theorem assures
the existence of an EDF
in which the energy is represented as a functional
only of the local density $n(\mathbf{r})$,
modification via the KS theory has been needed in its practical application
to many-fermion systems.
This fact is linked to the shell structure
that often emerges in many-fermion systems.
To illustrate this situation
and help to comprehend the path from the HK theorem to the KS theory,
we consider a simple system
under the 1-dimensional harmonic oscillator (h.o.) potential.

Let us take the Hamiltonian of the 1-dimensional h.o.,
\begin{equation}
  H = K + U\,;\quad K = -\frac{1}{2m}\,\frac{d^2}{dx^2}\,,~
  U(x) = \frac{m\omega^2}{2}x^2\,.
  \label{eq:H_HO}
\end{equation}
This Hamiltonian gives energy levels
$\epsilon_\nu=(\nu+\frac{1}{2})\omega$~$(\nu=0,1,2,\cdots)$,
which can be regarded as orbits.
We assume that each level has $\Omega$-fold degeneracy\footnote{
  This is accomplished when the particle has the spin $(\Omega-1)/2$.
  It can also be connected to a higher-dimensional spherical system
  by identifying $x$ as the radial coordinate
  and averaging out the angular and spin degrees of freedom.},
and $N$ non-interacting fermions occupy the orbits.
All particles occupy the $\nu=0$ orbit at the g.s. for $0<N\leq \Omega$,
$(N-\Omega)$ particles occupy the $\nu=1$ orbit for $\Omega<N\leq 2\Omega$,
and so forth,
providing the g.s. energy and density,
\begin{equation}\begin{split}
  & E^{(0)}_N = \left\{\begin{array}{ll}
  \frac{1}{2}N\omega &(0<N\leq \Omega) \\
  \big[\frac{1}{2}\Omega+\frac{3}{2}(N-\Omega)\big]\omega &
  (\Omega<N\leq 2\Omega) \\
  \quad\vdots &
  \end{array}\right.\,,\\
  &\quad n^{(0)}_N(x) = \left\{\begin{array}{ll}
    {\displaystyle\sqrt{\frac{m\omega}{\pi}}\,N\,e^{-m\omega x^2}} &
    (0<N\leq \Omega) \\
    {\displaystyle\sqrt{\frac{m\omega}{\pi}}\,
      \big[\Omega+2(N-\Omega)m\omega x^2\big]\,
    e^{-m\omega x^2}} &(\Omega<N\leq 2\Omega) \\
  \quad\vdots &
  \end{array}\right.\,.
\end{split} \label{eq:HO_exact}\end{equation}
This system is treated by the EDF under the HK theorem with $\mathbf{Q}=n(x)$.

The minimization of Eq.~\eqref{eq:EHK} can be substituted
by the Legendre transformation around the minimum.
The main task is constructing the kinetic part of the EDF.
The kinetic part of the EDF is desirable not to depend on $U$,
which is conjugate to $\mathbf{Q}\,(=n(x))$.
For $0<N\leq \Omega$, the following EDF~\cite{Stringari-Treiner_1987} is exact,
\begin{equation}
  E^\mathrm{HK}[n]
  = \int dx\,\bigg[\frac{1}{8m}\,\frac{\big(n'(x)\big)^2}{n(x)}
    + U(x)\,n(x)\bigg]\,.
  \label{eq:EDF_HO_0}
\end{equation}
Indeed, the variational equation of $E^\mathrm{HK}[n]$
with respect to $n(x)$ under the constraint $\int dx\,n(x) = N$,
\begin{equation}
  \delta\tilde{E}^\mathrm{HK}[n] =0\,;\quad
  \tilde{E}^\mathrm{HK}[n] := E^\mathrm{HK}[n]
  - \mu\bigg[\,\int dx\,n(x) - N\bigg]\,,
  \label{eq:var_HO-1}
\end{equation}
yields
\begin{equation}
  \frac{1}{4m}\bigg[\frac{n''(x)}{n(x)}
    -\frac{1}{2}\Big\{\frac{n'(x)}{n(x)}\Big\}^2\bigg] - U(x) + \mu = 0\,,
  \label{eq:var_HO-2}
\end{equation}
coinciding with Eq.~\eqref{eq:HO_exact}
when $U(x)=m\omega^2x^2/2$ and $\mu=\epsilon_0$.

The EDF of Eq.~\eqref{eq:EDF_HO_0} is not appropriate as it is
in $N>\Omega$.
The exact energy and density of Eq.~\eqref{eq:HO_exact}
for $\Omega<N\leq 2\Omega$
do not satisfy Eq.~\eqref{eq:var_HO-1}.
On the other hand,
it is always possible to obtain an exact functional
around a fixed $N=N_0$
in terms of $\mathit{\Delta}n(x):=n(x)-n^{(0)}_{N_0}(x)$,
\begin{equation}
  E^\mathrm{HK}[\mathit{\Delta}n] - E^{(0)}_{N_0}
  = \int dx\,\bigg[\frac{1}{8m}\,
    \frac{\big(\mathit{\Delta}n'(x)\big)^2}{\mathit{\Delta}n(x)}
    + U(x)\,\mathit{\Delta}n(x)\bigg]\,,
\label{eq:EDF_HO_1}\end{equation}
because $\mathit{\Delta}n(x)\propto|\varphi_{\nu_\mathrm{F}}(x)|^2$
($\nu_\mathrm{F}$ is the orbit just above $N_0$).
However, Eq.~\eqref{eq:EDF_HO_1} does not hold across $N=\Omega$;
the EDF of \eqref{eq:EDF_HO_1} is not correct if $N_0<\Omega<N$.
For Eq.~\eqref{eq:EDF_HO_1} to be valid for $N>\Omega$,
$n^{(0)}_{N_0}(x)$ and $E^{(0)}_{N_0}$ have to be redefined by shifting $N_0$.
This result provides evidence
that the EDF guaranteed by the HK theorem (Eq.~\eqref{eq:EHK})
is not regular everywhere.
This irregularity comes out because of the shell structure,
\textit{i.e.}, because the particles start occupying the $\nu=1$ orbit
for $N>\Omega$.
The source of the irregularity is the kinetic energy.

In contrast, the g.s. energy is immediately written
with the 1-body density matrix in the present non-interacting problem.
In the coordinate representation,
the density matrix is $\varrho(x,x')
:=\langle\Psi|\psi^\dagger(x')\,\psi(x)|\Psi\rangle$,
where $\psi(x)$ and $\psi^\dagger(x)$ are field operators.
The energy is exactly expressed as
\begin{equation}
  E[\varrho] = \int dx\,dx'\,\delta(x-x')\bigg[
    \frac{1}{2m}\,\frac{\partial^2}{\partial x\,\partial x'}\varrho(x,x')
    + U(x)\,\varrho(x,x')\bigg]\,.
\label{eq:EDF_HO_a}
\end{equation}
Identifying the kinetic term as the irregular part
and handling it as discussed in Sec.~\ref{sec:KS},
the EDF reaches
\begin{equation}\begin{split}
  E^\mathrm{KS}[\varrho;n] &= \frac{1}{2m}\int dx\,dx'\,\delta(x-x')\,
    \frac{\partial^2}{\partial x\,\partial x'}\varrho(x,x')
    + \int dx\,U(x)\,n(x) \\
  &= -\frac{1}{2m}\int dx\,dx'\,\delta''(x-x')\,\varrho(x,x')
    + \int dx\,U(x)\,n(x)\,,
\end{split}\label{eq:EDF_HO_b}
\end{equation}
equivalent to Eq.~\eqref{eq:EDF_HO_a}.
The EDF of \eqref{eq:EDF_HO_b} is global
and differentiable with respect to $(\varrho,n)$.

\bibliography{DFT}

\end{document}